\documentstyle[12pt,emlines,epsfig]{article}\textwidth=165mm
\makeatletter\@addtoreset{equation}{section}
\makeatother\def\theequation{\thesection.\arabic{equation}}
\begin{document}

\newpage
\pagenumbering{roman}

\title{ 
\bf
Cooling of Particle Beams in Storage Rings}\date{}
\author{E.G.~Bessonov, \\ Lebedev Physical Institute RAS, Moscow,
Russia}\maketitle

                       \begin{abstract}
Old and new cooling methods are discussed in reference to $e^{\pm}$,
ion and $\mu^{\pm}$ beams. \end{abstract}

                     \tableofcontents

\newpage
\pagenumbering{arabic}
\setcounter{page}{1}
                     \section{Introduction}

Different cooling methods were suggested to decrease the emittances of
charged particle beams in storage rings. Among them methods based on
synchrotron radiation damping \cite{bohm}-\cite{kolom-leb}, electron
cooling \cite{budker}, laser cooling in traps \cite{wineland} and in
storage rings \cite{channel}, \cite{shroder,hangst1}, ionization cooling
\cite{robinson}, \cite{oneil}-\cite{skrin}, cooing of ions through the
inelastic intrabeam scattering \cite{rubbia}, \cite{grim},
\cite{lauer}, and stochastic cooling \cite{vanderm}. The majority of
cooling methods are based on a friction of particles in external
electromagnetic fields or in media when the Liouville's theorem does
not work. Only stochastic method of cooling is not based on a friction.
It consists in the individual observation of particles and the action
of external control fields introduced in the storage ring for the
correction of particle trajectories.

The friction and corresponding energy losses are determined by the next
processes:

1) the spontaneous incoherent emission of the electromagnetic radiation
in external fields produced by bending magnets, undulators/wigglers,
laser beams etc.

2) ionization and excitation of atoms of a target at rest installed on
the orbit of the storage ring,

3) the transfer of the kinetic energy from particles of a being cooled
beam to particles of a co-propagating cold beam of $e^{-}$, $e^{+}$ or
ions in the process of the elastic scattering \cite{skrin},

4) excitation of being cooled ions and emission of photons by these
ions through the inelastic intrabeam scattering \cite{rubbia},
\cite{grim}, \cite{lauer},

5) the $e^{\pm}$ pair production by photons of a laser beam in fields
of being cooled ions \cite{bw}.

A friction originating from a media or in the process of emission
(scattering) of photons by charged particles in external fields leads,
under definite conditions, to a damping of amplitudes of both betatron
and phase oscillations of these particles when they are captured in
buckets of storage rings, i.e. there occurs a three-dimensional cooling
of particle beams\footnote{This conclusion is valid for all possible
types of friction processes and was emphasized in some pioneer papers
on synchrotron radiation and ionizing damping/cooling (see, e.g.
\cite{robinson, kolom}).}.  In this case particles of a beam loose
theirs momentum. At that the friction force is parallel to the particle
velocity, and therefore the momentum losses include both the transverse
and longitudinal ones.  Longitudinal momentum losses are compensated by
a radio frequency accelerating system of the storage ring. Meanwhile
the longitudinal momentum of a particle tends to a certain equilibrium.
This is because the rate of the momentum loss of the particle is
higher/less than the equilibrium value when this momentum is
higher/less than the equilibrium value. The transverse vertical
momentum of particles disappears irreversibly. Such a way the
compression of phase-space density for a given ensemble of particles
takes place.

The transverse radial and longitudinal particle oscillations are
dispersion coupled through energy losses \cite{kolom-leb}\footnote
{Owing to this coupling radial betatron oscillations can have negative
damping rate.}. Their damping rates can be corrected or redistributed
by insertion devices\footnote{Undulators, laser beams, wedge shaped
material targets having in general case non-homogeneous field
strengths or density in the radial direction at the position of being
cooled particle beam .} placed at straight sections of storage rings to
introduce an additional friction of particles, and by a correction of
the ring lattice parameters. The damping rates of both transverse and
longitudinal oscillations can be redistributed by coupling transverse
and longitudinal particle oscillations near betatron and
synchro-betatron resonances \cite{robinson}.

In the case of emission of particles in a laser wave the radiation
friction force appears together with the laser wave pressure force.
Friction force is determined by self-fields of the accelerated
particles and the pressure force is determined by the electromagnetic
fields of the laser wave (average value of the vector product of the
particle velocity and the magnetic field strength of the laser wave
which is not equal to zero when the self-force is taken into account)
\cite{zel'dovich}. The total averaged force can be directed both in the
same direction, or in the transverse and the opposite directions
relative to the average particle velocity. It depends on the direction
of the particle velocity relative to the wave propagation. We will
consider the relativistic case $\gamma \gg 1$ and conditions
corresponding to the interaction angle between vectors of the particle
velocity and the direction of the laser beam propagation $\theta_{int}
\gg 1/\gamma$, where $\gamma = \varepsilon/Mc^2$ is the relativistic
factor of the particle, $M$ particle rest mass, and $\varepsilon $ the
energy of the particle\footnote{In the case $\theta_{int} < 1/\gamma$
the particles will be accelerated by the light pressure force.}. In
this case both the radiation friction force and the pressure force are
directed opposite to the particle velocity.

The process of cooling of particle beams based on friction forces has a
classical nature and can be described in the framework of classical
electrodynamics. However, such data as atomic and nuclear levels,
oscillator strengths, degeneracy parameter, and so on, which determine
the corresponding laser wavelengths, cross-sections of particle
interactions and friction forces should be used from quantum mechanics.
At the same time, the excitation of longitudinal and transverse
oscillations of particles in storage rings has a quantum nature. In
this case the quasi-classical description of excitation of these
oscillations can be used \cite{kolom-leb}.

In the ordinary three-dimensional cooling, the particle beams are
cooled under conditions when all particles interact with the external
fields or media independent of their energy and amplitudes of betatron
oscillations. Insertion devices/targets in this case (wigglers, laser
beams, material media) overlap the particle beam and are motionless. In
this case the difference in the rates of energy losses of particles of
the beam having maximum and minimum energies is not high. That is why
the cooling time of the particle beam determined by this difference is
high and equal to the time interval, at which particle energy losses in
the target are equal to about the two-fold initial energy of the
particle\footnote{ This conclusion is valid for damping times both in
the longitudinal and transverse planes.}. At that the particle energy
must be recovered by the RF system of the storage ring.

Cooling of particle beams can be enhanced. In this case the damping
time of the particle beams in the longitudinal phase space can be
reduced essentially (in the energy of particle over the energy spread
of the particle beam ratio) if we will use selective interactions of
particles of the beam and targets. For this purpose we have to choose
such targets which interact with particles having definite energies or
amplitudes of betatron oscillations and do not interact with another
particles of the same beam.  For example, target can interact with
particles of the energy higher then minimum energy of the beam and do
not interact with particles of minimum and lesser energy. In this case
the rate of the energy loss of particles is not increased but the
difference in the rates of losses of particles of the beam having
maximum and minimum energies will be increased essentially and all
particles will be gathered at the minimum energy in a short time (equal
to the time interval, at which a particle looses the energy equal to
the initial energy spread of the particle beam).

In this paper some peculiarities of ordinary and enhanced cooling
methods based on a friction are discussed in reference to electron, ion
and muon storage rings. The main characteristics of cooling methods
(damping time, equilibrium emittance of the being cooled beam) will be
presented. Cooling of electron and ion beams in linear accelerators
will be presented as well.

\section{Three-dimensional radiative cooling of particle \\beams in
storage rings by laser beams}

In the method of the ordinary three-dimensional radiative laser cooling
of particle beams in buckets of storage rings, a laser beam overlaps a
particle beam at a part of its orbit. We shall consider a cooling
configuration where the laser beam is colliding head-on with a particle
beam in a dispersion-free straight section of a storage ring.  Both the
laser beam and the particle beam are focused to a waist at the center
of the straight section of the storage ring. In limits of this region
the particle beam is affected by a friction force through scattering of
laser photons. Particles lose their energy mainly in the process of
backward Compton or Rayleigh scattering of laser photons. The
accelerating fields of the radiofrequency systems of storage rings
compensate the radiative losses of the particle energy.  We assume that
the incident laser beam has a uniform spectral intensity $I _{\omega} =
dI_L/d\omega = I_L/\Delta \omega _L$ in the frequency interval $\Delta
\omega _L$ centered around $\omega _L$, where $I _L$ is the total
intensity (power per unit area).

In the case of an ion cooling, the electronic transitions of the not
fully stripped ions or nuclear transitions and broadband laser beam
have to be used. If the ion beam has an angular spread $\Delta \psi$,
around $\psi = 0$, and a relative energy spread $\Delta \gamma$, around
average $\overline \gamma$, the full bandwidth required for the
incoming laser to interact with all ions simultaneously (to shorten the
damping time of radial betatron oscillations) is $(\Delta \omega
/\omega)_L = (\Delta \psi)^2/4 + \Delta \gamma /\overline \gamma,$
where $\omega _L = \omega _{tr}/ \overline {\gamma }( 1 + \overline {
\beta _z})$; $\omega _{tr}$, the resonant transition energy in the rest
frame of the ion; $\beta _z = v _z/c$; and $v_z$, the longitudinal
component of the vector of the particle velocity. In the case of
radiative cooling of electron or fully stripped ion beams (Compton
scattering) a monochromatic laser beam can be used.

The physics of damping in this method is similar to a synchrotron
radiation damping originating from a particle emission in bending
magnets of storage rings. The difference is in the appearance of other
regions, where the emission of photons takes place, lattice parameters
of these regions, and in spectral distributions of the emitted
(scattered) photons.

Equilibrium emittances of particle beams are determined by a product of
damping times and the rates of excitation of longitudinal or transverse
oscillations of particles in the storage rings. The rate of excitation
of particle oscillations is determined by hardness and power of the
emitted radiation and by the lattice features of the storage ring such
as its global parameter "momentum compaction factor $\alpha$" and its
local parameter "dispersion function" in the regions where the particle
emits the radiation \cite{kolom-leb}. By analogy with synchrotron
radiation damping, in order to shorten a bunch length in a storage
ring, one should reduce $\alpha \to 0$ by manipulating with the ring
optics \cite{robinson2, murphy}. To shorten the transverse radial
emittance one should use long dispersion-free straight sections filled
with strong wigglers or laser beams (to produce fast damping without
additional excitation of betatron oscillations). In the case of
cooling of electron beams by lasers, the lattice of the storage ring
must have large-radius arcs with strongly focused FODO to produce low
quanta excitation by synchrotron radiation in bending magnets of the
ring \cite{hutton}-\cite{emery}.

\subsection{Three-dimensional radiative cooling of ion beams in
storage\\ rings by broadband lasers}

In the three-dimensional laser cooling, the ion beams are cooled under
conditions of Rayleigh scattering of laser photons when all ions
interact with the homogeneous laser beam independent of theirs energy
and position \cite{idea}-\cite{pac}. In this case, the average
cross-section of the photo-ion interactions, $\overline \sigma = \pi
f_{12}r_e\lambda _{tr}(\omega /\Delta \omega )_L$, is larger than the
Compton (Thompson) cross-section, $\sigma _T \simeq 8\pi r_e^2/3 \simeq
6.65 \cdot 10^{-25}$ cm$^2$, by about a factor of $(\lambda _{tr}/r_e)
(\omega /\Delta \omega )_L$, which is large, about $10^6 - 10^9$ for
many cases of the practical interest.  In the previous expressions,
value $f_{12}$ is the oscillator strength, $r_e = e^2/mc^2$ the
classical electron radius, $\lambda_{tr} = 2\pi c/\omega _{tr}$.

Assuming that photo-ion interaction takes place in dispersion-free
straight section, the damping time of horizontal betatron oscillation
$\tau _x$ is the same as the vertical oscillation $\tau _y$, because a
variation in the radiated energy due to a variation in the orbit
vanishes. The damping time of amplitudes of betatron and phase
($\tau_{\epsilon}$) oscillations of ions is

        \begin{equation}   
        \tau_x = \tau _y = {\tau
        _{\epsilon}\over(1+D)} = {2\varepsilon \over P},
        \end{equation}
where $P = f \Delta N _{int} \varepsilon _{loss}$ is the average power
of the electromagnetic radiation emitted (scattered) by the ion; $f$,
the frequency of the ion beam revolution in the storage ring, $\Delta N
_{int} = (1 + \beta)I _{sat}l_{int} \overline \sigma D/ c\hbar \omega
_L(1 + D)$, the number of ion interactions with the laser beam photons
per one ion-laser beam collision, $l _{int}$ the length of the
interaction region of the laser and ion beams; $D = I _L/I_{sat}$, the
saturation parameter; $I_{sat} = [g_1/4(g_1 + g_2)](\hbar \omega
_{tr}^4/ \pi^2c^2\gamma \overline {\gamma})(\Delta \omega /\omega )_L$,
the saturation intensity, $g_1 (g_2)$, the degeneracy factor of the
state 1(2); $\varepsilon _{loss} = \hbar \omega _{tr}\gamma = (1 +
\beta)\hbar \omega _L\gamma \overline {\gamma} $ the average loss of
the ion energy per one event of ion-photon interaction.

The expression $\tau _{\epsilon}$ in Eq(2.1) is specific to the
assumption that the spectral intensity of the laser beam $I _{\omega}
(\omega, x,y)$ is constant inside its bandwidth and inside the area of
the laser beam occupied by the being cooled ion beam\footnote{The
presence of a derivative $\partial I _{\omega}/\partial {\omega}$ will
lead to another value $\tau _{\epsilon }$ \cite{prl}. The
longitudinal-radial coupling arising in non-zero dispersion straight
sections of the storage rings leads to a redistribution of the
longitudinal and radial damping times when the radial gradient of the
laser beam intensity $\partial I _{\omega}/\partial {x}$ is introduced
\cite[c]{idea}. Experiments \cite{lauer} confirm this observation. The
same idea is in the three-dimensional scheme of the ionizing cooling of
muon beams through dispersion coupling in the wedge-shaped material
target. It was discussed in the paper of D.V.Neuffer \cite{neuffer}.}.
Moreover, we assume that the length of the ion decay $l_{dec} = c ^2g
_2\beta \gamma/ 2g_1 f_{12} r_e \omega _{tr}^2 = g_2 \beta \gamma
\lambda _{tr} ^2/8\pi ^2g_1 f_{12} r_e$ is much less than the length of
the dispersion-free straight section \cite{idea}-\cite{pac}.  The
length of the ion decay is determined by the spontaneous decay time
$\tau _{sp} = 1/\Gamma _{21}$, where $\Gamma _{21} = 2g _1 f_{12}
r_e\omega _{tr}^2/g_2 c$ is the probability of the spontaneous photon
emission of the excited ion or the natural linewidth $\Delta {\omega}
_{nat}$.  Usually the relative natural linewidth ($\Delta \omega/
\omega)_{nat} = 4\pi f_{12}(g_1/g_2)(r_e/\lambda _{tr})$ is less than
the line width of a laser beam $(\Delta \omega /\omega )_L$ necessary
for a three-dimensional ion cooling by broadband laser beams, and is
determined by the energy and angular spreads of ion beams. Otherwise
the monochromatic laser beams can be effectively used for the same
purpose.

The quantum nature of the laser-photon scattering provides excitation
of betatron and phase oscillations of ions. The calculation of
equilibrium amplitudes is similar to the case of ordinary electron
storage rings, except that the spectral-angular probability
distribution of the scattered photons here is given by that of the
undulator radiation. We have found that the relative r.m.s. energy
spread of the ion beam at equilibrium is given by \cite{prl}

        \begin{equation}   
        {\sigma_{\varepsilon}\over \varepsilon} = \sqrt{1.4(1 + D)\hbar
        \omega _{tr}/Mc^2}. \end{equation}

In the present case, where the interaction takes place in a
dispersion-free straight section, the excitation of both the horizontal
and vertical betatron motions is due to the fact that
the propagation direction of emitted photons is not exactly
parallel to the vector of the ion momentum. The r.m.s. equilibrium
horizontal ion beam emittance is found to be

        \begin{equation}   
        \epsilon_{x} = {3\over 20}{\hbar \omega _{tr}\over \gamma ^2
        Mc^2} <\beta_x>. \end{equation}

In (2.3), $<\beta_x>$ is the average horizontal beta function in the
interaction region. The equilibrium vertical emittance is obtained by
replacing $<\beta _x>$ by $<\beta _y>$. The r.m.s. transverse size of
the ion beam at the waist $\sigma _x = \sqrt{\beta _x \cdot
\epsilon_{x}}$.

In practice the three-dimensional method of laser cooling of ion beams
have to realize after one-dimensional one (see Section 3). This will
permit to cool the ion beam in the longitudinal plane for a short time
using low power monochromatic laser beam in the first case and then to
shorten the bandwidth and power of the laser beam in the second one.
After that the one dimensional cooling can be used again to decrease
the longitudinal emittance of the being cooled ion beam\footnote{The
energy spread of the being cooled beam is minimum in the case of
one-dimensional laser cooling.}.

The three-dimensional laser cooling of ion beams can be realized by a
monochromatic laser beam with accelerating fields of radio frequency
cavities as well. The nature of the transverse cooling by the
monochromatic laser beam in the three-dimensional method of cooling
does not differ from the case of the broadband laser cooling. A
difference is in conditions of interaction of ion and laser beams. In
the case of a broadband laser beam all ions interact with the laser
beam independent of their energies and amplitudes of betatron
oscillations. In the case of a monochromatic laser beam every ion
interacts with the laser beam only a part of time, when it passes the
ion resonance energy in the process of phase oscillations in a bucket
of a storage ring. That is why it has greater damping time at the same
saturation parameter as in the case the broadband laser and the same
cooling configuration (at that the power of the broadband laser is
higher). Smaller value of the transverse damping time is the advantage
of the broadband laser cooling in the case of a three-dimensional laser
cooling of ion beams.

Experimentally, the version of longitudinal cooling of a bunched
non-relativistic beam of $^{24}Mg^{+}$ ions (kinetic energy $\sim$ 100
keV) was observed first in the storage ring ASTRID \cite{hangst}. The
monochromatic laser beam co-propagated with the ion beam (conditions of
the ion acceleration) at scanning of its frequency and using the
accelerating system of the storage ring. At such cooling, some degree
of the radiative transverse cooling could be observed.

\subsection{Three-dimensional radiative cooling of electron beams in
\\storage rings by lasers}

A three-dimensional cooling of electron and proton beams based on the
backward Compton scattering of laser photons in the dispersion-free
straight sections of the storage rings can be used \cite{idea, prl,
kim, zhirong}. In this case we can use the expressions (2.1) - (2.3) if
we replace the values $\overline \sigma \to \sigma _T$, $I_{sat}D =
I_L$ and accept $D = 1$\footnote{The difference in the physics of
cooling of electron and ion beams is in the dependence of the average
hardness $\varepsilon _{loss}$ and power $P$ of scattered radiation on
the energy $\gamma$. The damping time of the longitudinal oscillations
and equilibrium energy spread of a particle beam are determined by the
derivative $\partial P/ \partial \gamma$. The powers emitted by
electron and ion are: $P _e \sim \gamma ^2$, $P _{ion} \sim \hbar
\omega \gamma$ ($\omega _{tr} = const$, ion of higher energy scatter
laser photons of lesser energy!). Because of this difference we will
have correct result for the case of cooling of electron beams if we
will choose $D = 1$ in (2.1), (2.2) \cite{prl}.}. The method is
identical to that suggested in papers \cite{hutton}-\cite{emery} where
magnetic wigglers were used instead of laser beams\footnote{
Electromagnetic waves can be considered as objects which belong to the
type of undulators/wigglers \cite{bes78}.  Wigglers with high
deflecting parameters can change the lattice parameters of the storage
ring.}. In this case the excitation of radial betatron oscillations
will take place only through the emission of photons of synchrotron
radiation from bending magnets of the ring where the dispersion
function differ from zero. At the same time, the rate of the betatron
oscillation damping  will be determined by total power emitted both in
the form of synchrotron radiation and Compton scattering of laser
photons.  The electromagnetic radiation emitted by electrons in the
process of Compton scattering can lead to a significant shortening of
damping time and hence equilibrium emittance of stored beams if the
power of scattered radiation will be much higher than the power of the
synchrotron radiation.

The method of the radiative cooling considered in this section is not
optimal. The damping time and the emittance of particle beams can be
shortened significantly by using a selective interaction of particles
and laser beams. Below we shall consider one- and two-dimensional
enhanced cooling schemes based on selective interactions of particle
beams and targets.

\section{Enhanced cooling of particle beams}

In the case of ordinary longitudinal cooling all particles of a beam
loose their energy in storage rings such a way that the ratio of a
difference in rates of the energy loss of particles having maximum and
minimum energies is small. The small difference in rates in comparison
with the rates of the energy loss of particles leads to a small
relative velocity of bringing closer of their energies. That is why the
damping time of longitudinal emittance of the particle beam is high. It
is determined by about two-fold loss of the initial particle energy in
the external fields or targets under conditions of recovering of the
energy in the radio frequency system of the storage ring.

Introduction of special damping magnets (wigglers) or wedge-shaped
targets can lead to decreasing of damping time of particles in the
longitudinal phase space. But such insertion devices simultaneously,
according to Robinson's damping criterion, lead to increase of damping
time of the particles in the transverse radial phase space or to
untidamping \cite{kolom-leb}. This is because of the Robinson's theorem
is valid for the case of linear systems, when target overlap the
particle beam completely and when longitudinal and transverse betatron
oscillations are dispersion coupled.

The damping time of the particle beam in the longitudinal phase space
can be reduced essentially (in the energy over the energy spread ratio)
if we will use selective interactions of a particle beam and a target
when only a part of particles of the beam interact with the target. In
this case we have to choose such targets and conditions of interaction,
when targets interact with particles having definite energies or
amplitudes of betatron oscillation and do not interact with another
particles of the same beam. For example, we can switch on interaction
between target and particles of the beam having maximum energy first,
extend in time continuously the interaction with particles of lesser
energy and switch off the interaction at the energy equal to minimum
energy of particles in the beam\footnote{At that the velocity of
extending of the interaction with particles must be higher then the
maximum velocity of shrinking of the instantaneous orbits of particles
under interaction with the target.}. Then we can repeat this
manipulation and gather all particles of the beam at minimal energy.
Such a way we will realize the cooling of a particle beam in the
longitudinal plane. In this case the rate of the energy loss by a
particle must depend mainly on the position of its instantaneous orbit
and must not depend on the deviation of the particle from this orbit
because of the betatron oscillations. The damping time of the particle
beam will be determined by the condition in which a particle having
maximum energy will loose the energy equal to the energy spread of the
being cooled beam.

The selectivity can be achieved in different ways. Among them note a
method based on a resonant and energy threshold interaction ($e^{\pm}$
pair production, interaction of ions with a broadband laser beam having
sharp frequency edges), a time dependent degree of overlapping in the
radial direction of a being cooled particle beam and a moving target.

We can see a similarity between enhanced longitudinal and stochastic
cooling of particle beams in the individual manipulations in turns with
parts of particle beams which are differ by definite parameters
(energies, amplitudes of betatron oscillations) in the first case and
individual particles in the second case. However this is only
similarity. The reason of the enhanced cooling is in a friction and the
selective interaction of particles and transversely moving target (the
degree of overlapping of the target and particle beam depends on time,
particles of the energy lesser then minimum one does not interact with
the target, the duration of interaction is proportional to the energy
deviation of the particle from the minimum initial energy of particles
in the beam). The enhancing effect is possible only in the case when
the cooled beam occupies a small lauer in the phase space. It is
proportional to the ratio of the minimum energy of the beam to its
energy spread. The enhanced cooling would be absent when the particles
of the beam where distributed in the range of energies from zero to
maximum one. Below we will discuss some possible enhanced schemes of
cooling.

\subsection{One-dimensional laser cooling of ion beams in storage
rings}

A typical version of one-dimensional laser cooling of ion beams is
based on the resonant interaction of unbunched ion beam and homogeneous
monochromatic laser beam overlapping the ion one in the transverse
direction. The initial frequency of the laser beam (photon energy) in
this version of cooling is chosen so that photons interact first with
the most high energy ions. Then the frequency is scanning (frequency
chirp) in the high frequency direction, and ions of a lower energy
begin to interact with the laser beam and decrease their energy. The
scanning of the laser frequency is stopped when all ions are gathered
at the minimum energy of ions in the beam.

The resonance ion energy $\varepsilon _r = Mc^2 \gamma _r$, where
$\gamma _r = [1 + (\omega _{tr}/ \omega _L)^2]/2(\omega _{tr}/\omega
_L)$ is the resonance relativistic factor. The initial resonance energy
is above the equilibrium energy of the storage ring, and is higher than
that corresponding to the maximum ion energy in the beam. Minimum ion
energy must be higher than the equilibrium one as well (the
accelerating RF electric field strength will be switched on after
cooling). The rate of scanning must correspond to the condition $\dot
\varepsilon _r = d\varepsilon _r/dt < P$, where $P$ corresponds to the
power of radiation scattered by an ion at resonance conditions.

The cooling time is equal to the time interval, at which the particle
with maximum initial energy emits the electromagnetic radiation energy
equal to the initial energy spread $\sigma _{\varepsilon \,in}$ of the
ion beam, and the energy spread of the cooled beam is determined by
either the average energy of the scattered photons $\varepsilon
_{loss}$ or the natural line width of the laser beam:

        \begin{equation}   
        {\tau _{\epsilon}} = {\sigma _{\varepsilon \,in}\over P},
        \hskip 20mm
        {\sigma _{\varepsilon } \over \varepsilon } = max\left(
        {\varepsilon _{loss} \over Mc^2 \gamma }={\hbar \omega _{tr}
        \over M c^2}, \hskip 5mm ({\Delta \omega \over
        \omega})_{nat}\right).  \end{equation}

This time is $\sim \varepsilon /\sigma _{\varepsilon \,in} \sim 10^3
\div 10^4$ times lower than that in the case of the three-dimensional
cooling. This is the consequence of the selective resonance interaction
of photon and ion beams in the one-dimensional method when ions of
the energy higher than minimum initial energy of the beam interact with
the laser beam.

The considered method is one of the possible one-dimensional ion
cooling methods\footnote{In another version of cooling, the laser
frequency can be constant, and the acceleration of ions in the
direction of given resonance energy can be produced by eddy electric
fields of a linear induction accelerator or by phase displacement
mechanism. The use of an induction accelerator is not efficient in high
energy ($\varepsilon > 1$ Gev/nucleon) storage rings.}. The first
similar method was used for cooling of non-relativistic ion beams
\cite{shroder}, \cite{hangst1}, \cite{petrich}\footnote{Two laser beams
of different frequencies, co- and counter-propagating with the ion
beam, can be used in the non-relativistic and moderate relativistic
case. In the coordinate system connected with the ion beam the
frequencies of laser waves can be equal and form a standing wave at the
resonance energy \cite{bond}.}. Relativistic version of such a method
was developed in \cite{habs}.

One-dimensional laser cooling of bunched ion beams by monochromatic
laser beam is possible with accelerating fields of radiofrequency
cavities (see section 2.1) \cite{hangst}. The broadband laser beam with
a sharp low frequency edge can be used as well. In this case the edge
frequency must have such a value that only ions with energies above the
equilibrium one can be excited \cite{besabstr}. To gather the cooled
ion beam into short bunches the radiofrequency accelerating cavity
should be switched on adiabatically.

\subsection{Two-dimensional cooling of particle beams in storage rings}

One-dimensional laser cooling is highly efficient in the longitudinal
direction, but rather difficult in the transverse direction unless a
special longitudinal-radial coupling mechanism is applied
(synchro-betatron resonance \cite{robinson, sessler, idea}, dispersion
coupling \cite{oneil, lauer}). Moreover this is the resonance method of
cooling. It can be applied to cooling of only complicated ions. In
papers \cite{idea}-\cite{pac} a three-dimensional radiative ion cooling
method is proposed and considered above.  Nevertheless, the quest for
new more efficient enhanced three-dimensional cooling methods remains
vital for cooling of electrons, protons, muons and both not fully
stripped and fully stripped high current ion beams. Below we will
discuss a two-dimensional method of cooling when alternative targets
and selective interactions are used\footnote {Coupling resonance of
betatron oscillations permits cooling of a particle beam in three
dimensions.}.

\begin{figure}[hbt]
\centerline{\leavevmode\epsfxsize=130mm\epsfbox[122 391 512
730]{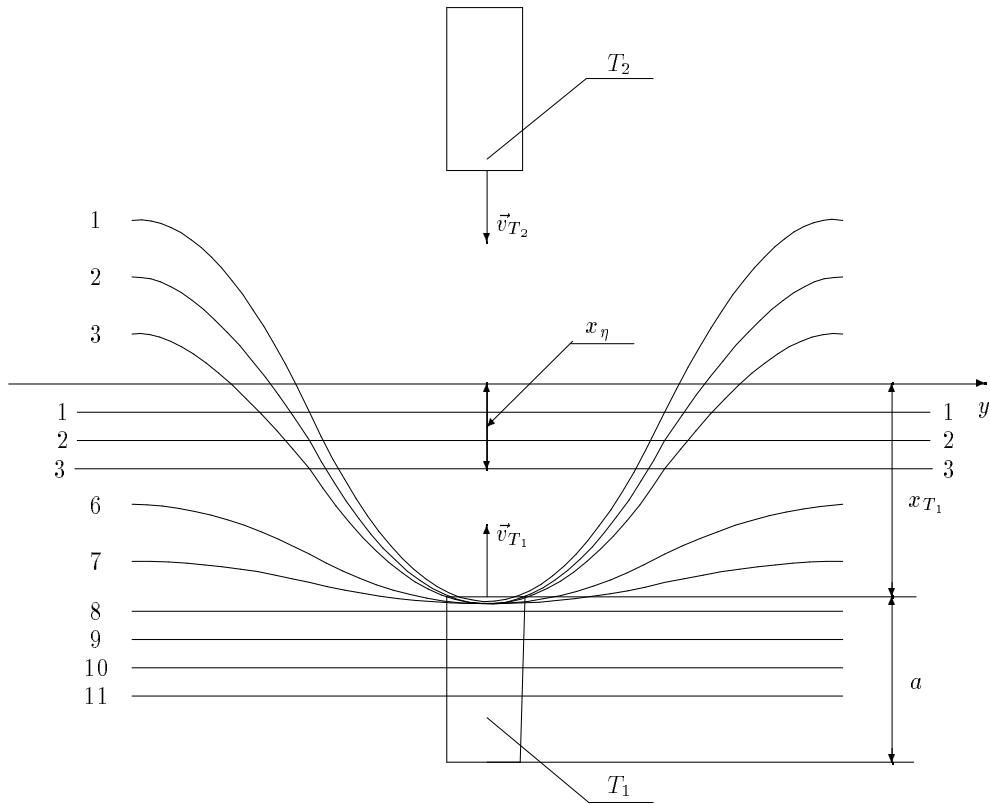}}
\caption{\small \it The scheme of the two-dimensional ion cooling.The
axis $y$ is the equilibrium orbit of the storage ring, 1-1, 2-2 ... the
location of the instantaneous ion orbit after 0,1,2 ...  events of the
ion energy loss, $T_1$and $T_2$ are the targets moving in turn with the
velocity $\vec v_{T_{1,2}}$ from outside to an equilibrium orbit.}
\end{figure}

We consider the process of change of amplitudes of betatron
oscillations $A$ and positions of instantaneous orbits $x _{\eta}$ in
the process of the energy loss of particles in targets when the RF
accelerating system of the storage ring is switched off \cite{bkw}. The
internal target $T_1$ at the first stage and external target $T_2$ of
constant thickness at the second one will be moved in the transverse
directions in turn (see Fig.1). They can be located in different
positions along the way of the particle beam in the storage ring. Every
stage of cooling can be used in combination with another schemes of
enhanced cooling or with schemes of the emittance exchange in the
longitudinal and transverse planes.

The velocity of a particle instantaneous orbit $\dot x_{\eta}$ depends
on the distance $x _{T _{1,\,2}} - x _{\eta}$ between the target and
the instantaneous orbit, and on the amplitude of particle oscillations.
When the instantaneous orbit of a particle enters the target at the
depth higher than the amplitude of the particle oscillations then its
velocity $\dot x_{\eta \, in}$ is maximum one by the value and
negative. The velocity $x _{\eta \, in}$ is given by the material
target thickness or the intensity and the length of the interaction
region of the laser target.

In the general case particles do not interact with the target every
turn. That is why the velocity $\dot x_{\eta} = \dot x_{\eta \,
in}\cdot W$, where $W < 1$ is the probability of a particle to cross
the target. The probability $W$ is determined by the ratio of the
period of betatron oscillations of a particle to a part of the period
which is determined by conditions: 1) the deviation of the particle
from the instantaneous orbit is greater than the distance between the
orbit and the target ($| x _{T _{1 \,2}} - x_{\eta}| \leq |x _{0}^{'}|
\leq A$), 2) the deviation is directed to the target. It can be
presented in the form $W = \Delta \varphi _{1, \,2}/\pi$, where $\Delta
\varphi _{1} = \pi - arccos \xi _{1}$, $\Delta \varphi _{2} = arccos
\xi _{2}$, $\xi _{1,\, 2} = (x _{T_{1, \,2}} - x _{\eta}) /A$, labels
1, 2 correspond to first and second targets used at first and second
stages of cooling accordingly.

In the approximation when random processes of excitation of betatron
oscillations of particles in a target of a storage ring can be
neglected the behavior of the amplitudes of betatron oscillations of
these particles is determined by the equation $\partial A^2 /\partial
x_{\eta} = - 2<x^{'}_0>$\, or \,$\partial A/\partial x _{\eta} =
-<x^{'} _0>/A$ (see Appendix A), where $<x^{'} _0>$ is the particle
deviation from the instantaneous orbit averaged through the range of
phases $2\Delta \varphi _{1, \,2}$ of betatron oscillations where the
particle cross the target\footnote{We suppose that particles loose
their energy in the target. When a complicated ion is excited in the
target, then we suppose that the length of decay of the excited ion is
much less then the length  of betatron oscillations of the particle.}.
The value $<x^{'} _0> = \pm A sinc \, \Delta \varphi _{1 \,2}$, where
$sinc \Delta \varphi _{1 \,2} = sin \Delta \varphi _{1 \,2}/\Delta
\varphi _{1 \,2}$, signs $+$ and $-$ are related to the first and
second stages of cooling. Thus the cooling processes at the first and
second stages are determined by the system of equations

       \begin{equation}
       {\partial A\over \partial x_{\eta}} = \pm sinc \Delta \varphi
       _{1,2}, \hskip 5mm {\partial x _{\eta}\over \partial t} = {\dot
       x_{\eta \, in} \over \pi}\Delta \varphi _{1,2}.
       \end{equation}
where $\Delta \varphi _{1} = \pi - arccos \xi _{1}$, $\Delta \varphi
_{2} = arccos \xi _{2}$, $\xi _{1,\, 2} = (x _{T_{1, \,2}} - x
_{\eta}) /A$.

These equations are valid both for one and for a set of thin foils of
the same total thickness, located at different azimuths and at definite
depths of storage rings. This system of equations permits to
investigate the main processes connected with cooling of particle
beams.

From the Eqs. (3.2) and the expression $\partial A/\partial x _{\eta} =
[\partial A/\partial t]/ [\partial x _{\eta}/\partial t]$ it follows:

       \begin{equation}
       {\partial A\over \partial t} = {\dot x _{\eta \,in}\over
       \pi}\sin\Delta \varphi _{2} = {\dot x _{\eta \,in}\over
       \pi}\sqrt {1 - \xi _{1,\,2}^2}.  \end{equation} 

Let the initial instantaneous particle orbits be distributed in a
region $\overline {x _{\eta}} \pm \sigma _{x, \varepsilon, 0}$ and
initial amplitudes of particle radial betatron oscillations $A _0$ be
distributed in a region $\sigma _{x,b,0}$ relative to their
instantaneous orbits, where $\overline {x _{\eta}}$ is the location of
the middle instantaneous orbit of particles of the beam; $\sigma _{x,
\varepsilon, 0}$, the mean-root square deviation of instantaneous
orbits from the middle one. The value $\sigma _{x, \varepsilon, 0}$ is
determined by the initial energy spread $\sigma _{\varepsilon, 0}$.
Suppose that the initial spread of amplitudes of betatron oscillations
of particles $\sigma _{x,b,0}$ is identical for all instantaneous
orbits of the beam. The velocity of the instantaneous orbit in a target
$\dot x _{\eta, in} < 0$, the velocity of the first target $v _{T _{1}}
> 0$ and the velocity of the second target  $v _{T _{2}} < 0$.  Below
we will use the relative velocities of targets $k _{1,2} = v _{T
_{1,2}}/\dot x _{\eta, in}$.  In our case $k _1 < 0$ and $k _2 > 0$.

   \begin{center} \bf First  stage of cooling \end{center}

At the first stage of the two-dimensional cooling, an internal target
$T_1$ (laser beam or a material medium target) is located at the orbit
region ($x _{T _1}$, $x_{T _1} - a$), where $a$ is the target width
(see Fig.\,1).  The internal edge of the target has the form of a flat
sharp boundary. At the initial step the target overlaps only a part of
the particle beam so that particles with largest initial amplitudes of
betatron oscillations interact with the target. The interaction takes
place only at the moment when the particle deviation caused by betatron
oscillations is directed toward the target and has approximately the
maximum initial amplitude $A _0 \simeq \sigma _{x,b,0}$.

In this case immediately after the interaction and loss of energy the
position of a particle and the direction of its momentum remain the
same, but the instantaneous orbit is displaced inward in the direction
of the target\footnote{We suppose that the friction leads to a decrease
in the momentum by the value only and neglect scattering of the
particle in the target for the time of the enhanced cooling}. The
radial coordinate of the instantaneous orbit and the amplitude of
betatron oscillations are decreased to the same value owing to the
dispersion coupling. After every interaction the position of the
instantaneous orbit will approach the target more and more and the
amplitude of betatron oscillations is coming smaller until it will
reach some small value when the instantaneous orbit will reach the edge
of the target.  Up to this moment the instantaneous orbit go in the
direction to the target, but the particle depth of dipping do not move
forward deeper into the target.

At the moment when the instantaneous orbit of the particle enter the
target the amplitude of the particle betatron oscillations is much
decreased.  The instantaneous orbit will continue its movement in the
target with constant velocity $\dot x_{\eta \, in}$ if the target is
homogeneous one, has constant thickness and when its depth of dipping
in the target is greater then the amplitude of particle betatron
oscillations\footnote{The amplitude will be increased in proportion to
the r.m.s. number of particle interactions with the target when random
processes of deviation of energy losses from an average value exist.
The amplitude increase here is slower than the amplitude reduction when
it changed proportionally to the number of interactions. We will
neglect this increase here.}.

The particle beam has a set of amplitudes of betatron oscillations and
instantaneous orbits. To cool the beam we must move the target in the
direction to the particle beam or move instantaneous orbits of the
particle beam in the direction to the target\footnote {A kick,
decreasing of the value of the magnetic field in bending magnets of the
storage ring, a phase displacement or eddy electric fields can be used
for this purpose.} at some velocity $v_{T_1}$ until the target will
reach the instantaneous orbit having maximum energies.  After all
particles of the beam will interact with the target, the latter must be
removed or the particle beam must be returned to the initial position
for a short time. After this, all particles will have small amplitudes
of betatron oscillations and increased energy spread.

Here we will estimate the possibilities of the transverse cooling of
particle beams at the first stage of cooling. Let the internal target
be displaced in the direction of the being cooled beam step by step
with a pitch $\Delta x _{T _1} = \sigma _{x,b,0}/M$, where $M \geq
2$ is the whole number. In this case at the first step of the first
stage of cooling the target overlaps a pitch $\Delta x _{T _1}$ of a
being cooled beam.  The instantaneous orbits of particles having
maximum amplitudes of betatron oscillations after the moment of
their entry into the target will be displaced in the direction of the
target with a velocity $\dot x _{\eta}$ and the amplitudes of betatron
oscillations will be reduced.  When $x _{\eta} - x _{T _1} \simeq A
_{f} \gg \Delta x _{T _1}$ then $\Delta \varphi _1 \simeq \sqrt
{2\Delta x _{T _1}/A _{f}} \ll 1$, $sinc \,\Delta \varphi_1 \simeq
1 - \Delta x _{T _1}/3A _{f} \simeq 1$ and according to (3.2) the
value $\partial A/\partial x _{\eta} \simeq 1$, $A(\xi _{1}) -
(x_{\eta} - x_{T _1}) \simeq \Delta x _{T _1}$ up to the time when the
instantaneous orbits approach the target to the distance $x _{\eta}
\simeq x _{T _1}$ and when final amplitude of betatron oscillations
$A _f \sim \Delta x _{T _1}$.  When the orbits enter the target
($\Delta \varphi_1 = \pi/2$) the value $dA/ dx_{\eta}$ is decreased by
$\pi/2$ times only.  Then the orbits of particles will be deepened into
the homogeneous target at the depth greater than their final amplitudes
of betatron oscillations $A _f$, i.e. when $\Delta \varphi _1 = 0$,
$\partial A/\partial x _{\eta} = 0$. After that the particles will
interact with the  target every turn without change of their
amplitudes. This means that amplitudes of betatron oscillations in the
first approximation will not tend to zero, but to some finite value $A
_f\sim \Delta x _{T _1} \sim \sigma _{x\,b\,0}/M$.

More accurate numerical calculations of the dependencies $A _{f}/A
_0$ and $A _{f}/ \Delta x _{T _1}$ based on the equation (B.5) for the
motionless target ($k _1 = 0$, $\xi _0 = -1 + \Delta x _{T_1}/A_0$.)
are presented in the Table 1.

\begin{table}[hbt]
Table 1
\vskip 3mm
\begin{tabular}{|l|l|l|l|l|l|l|}
\hline
$\Delta x _{T _1}/A _0$ & 0.000 & 0.010& 0.050 & 0.100 & 0.500 & 1.00 \\
\hline
$A _{f}/A _0$ & 0.000 & 0.024& 0.082 & 0.137 & 0.442 & 0.707 \\
\hline
$A _{f}/\Delta x _{T _1}$& ----- & 2.40& 1.640 & 1.37 & 0.884 &0.707\\
\hline
\end{tabular}
\end{table}
\vskip 5mm

The regime of the target movement step by step was considered for
illustration of the cooling mechanism. In practice we can move the
target uniformly with the velocity $v _{T _1}$. Detailed numerical
calculations of the first stage of the transverse cooling in this case
are presented in the Appendix B. According to these calculations $A
_f/A _0 = \sqrt {|k _1|/(|k _1| +1)}$.

The time of the target movement in the first stage of cooling is
$\Delta t _1 \simeq \sigma _{x, 0}/v _{T_1}$, where $\sigma _{x ,0} =
\sigma _{x,b,0} + \sigma _{x,\varepsilon,0}$ is the total initial
radial dimension of the particle beam. For this time the instantaneous
orbits of particles of a beam having minimum energy and maximum
amplitudes of betatron oscillations will pass the distance $\simeq
|\dot x _{\eta\, in}| \Delta t_1$. Hence the final radial dimension of
the beam determined by the final energy spread of the beam and the
total radial dimension of the beam will be increased to the value

         \begin{equation}
         \sigma _{x,f} \simeq  \sigma _{x,\varepsilon, f} = \sigma
         _{x,0} {1\over |k _1|}.
         \end{equation}   

Thus, at the first stage we have a high degree enhanced cooling of
particle beams in the transverse plane (3.4) and more high degree of
heating in the longitudinal one (B.6).

At the first stage of the two-dimensional transverse cooling of
particle beams it is desirable to use the straight section with
low-beta and high dispersion functions. In this case, less events of
the photon emission are required to cool the beam in the transverse
direction.  This is because the change of amplitudes of betatron
oscillations is the same as the change of positions of instantaneous
orbits of the particle.  Meanwhile, the spread of amplitudes of
betatron oscillations is small and the step between positions of
instantaneous orbits is high.

       \begin{center} \bf Second stage of cooling \end{center}

At the second stage of cooling an external target $T_2$ is moving with
a velocity $v_{T _2} < 0$ from outside of the working region of the
storage ring in the direction of a being cooled particle beam (or
instantaneous orbits of the particle beam are moved in the direction of
the target). The instantaneous orbits of particles will go in the same
direction with a velocity $\dot x _{\eta} \leq \dot x _{\eta, in}$
after the moment of their first interaction with the target. In this
case, the target will start to interact first with particles having the
largest amplitudes of betatron oscillations and the highest energies at
some moment $t _0$.  Then it will interact with particles of lesser
amplitudes and energies.  When the target will pass through the
instantaneous orbit of particles having zero amplitudes and minimum
initial energies then it must be removed to the initial position.

We will start from the estimation of the value of the transverse
heating.  According to (3.2) and (3.3) the value $\partial A/ \partial
t = (\dot x _{\eta \,in}/ \pi) \sqrt {1 - \xi _{1,\,2}^2} \leq |\dot x
_{\eta \,in}|/ \pi$ for the arbitrary time and the value $|\partial x
_{\eta}/ \partial t| \leq |\dot x _{\eta \,in}|/2$ when $x _{T_2} - x
_{\eta} \geq 0$ that is up to the time $t _{1/2} \simeq t_0 + A _0/|v
_{T_2} - \dot x _{\eta \, in}/2|$ when the target will reach the
instantaneous orbit.  For the time $t _{1/2} - t_0$ the target will
pass the way $\Delta x _{T_2} < 2A _0 k_{2}/(2k_{2} - 1)$, the
instantaneous orbit will pass the way $\Delta x _{\eta} < A _0/(2k _2 -
1)$ and the increase of the amplitude of betatron oscillations will be
$\Delta A < 2A _0k _2/\pi(k _2 - 1)$.  Specifically, when $v _{T_2} =
1.5 \dot x _{\eta \, in}$ the values $\Delta x _{T _2} = 1.5 A_0$,
$\Delta x _{\eta} < A_0/2$, $\Delta A _1< A_0/\pi$.

After the target passed the instantaneous orbit at a moment $t _c$ then
in a time interval $t _{c} - t _{1/2}$ the value $\dot x _{\eta}$ is
increased to the value $\dot x _{\eta \, in}$, and $\partial A/\partial
t$ is decreased to zero. At that the value $x _{\eta } < x _{T _2} + A$
and the amplitude of betatron oscillations will reach its final value
$A _f$. Specifically, when $v _{T_2} = 1.5 \dot x _{\eta \, in}$ then
$t _c - t _0 \simeq 2(t _{1/2} - t_0)$, and at this step the increase
of the amplitude of betatron oscillations will be $\Delta A _2 < A_0(1
+ 1/\pi)(1/\pi) \simeq 0.42A _0$. Finally we will have the amplitude $A
_f \simeq A _0 + \Delta A _1 + \Delta A _2 \simeq 1.74 A _0$.

Now we will estimate the behavior of the energy spread of the beam in
the second stage of cooling. Let, for the simplicity, that the initial
spread of positions of instantaneous orbits $\sigma _{x,\varepsilon,0}$
is much greater then the spread of the amplitudes of betatron
oscillations $\sigma _{x, b, 0}$ of the beam. In this case high energy
particles first and then particles with smaller energies will interact
with the moving target until the target will reach the instantaneous
orbit with the least energy. Then the target must be removed or the
particle beam instantaneous orbits must be returned to the initial
position for a short time.

The particles of the beam having maximum energy and zero amplitudes of
betatron oscillations will interact with the target during the time
$\Delta t _2 ^{'} \simeq \sigma _{x,\varepsilon,0}/v _{T _2}$. For this
time the instantaneous orbits of particles will pass the distance
$|\dot x _{\eta\, in}| \Delta t _2 {'} = k _2 ^{-1} \sigma
_{\varepsilon, b,0}$. At that particles having minimum energy and zero
amplitudes of betatron oscillations will stay at rest. Hence it follows
that the spread of instantaneous orbits of these particles will be
compressed to the value $\sigma _{x,\varepsilon,f} \simeq \sigma
_{x,\varepsilon,0} {(1 - k _2 ^{-1})}$. If we take into account the
fact that the behavior of the instantaneous orbit depends on the
initial amplitude of particle betatron oscillations then the total
radial dimension of the beam can be presented in the form $\sigma
_{x,\varepsilon,f} = \sigma _{x,\varepsilon,0} {(k_2 - 1)/ k _2} +
\Delta \sigma _{x,\varepsilon}$, which, according to the numerical
calculations produced in the Appendix B, can be represented in the form
($t > t _0 ^{'}$)

         \begin{equation}
         \sigma _{x,\varepsilon,f} \leq \cases { [{k_2 - 1\over k _2}
         {\sigma _{x,0} - A _{T _2} \over \sigma _{x,b,0}} - \xi
         _{2,st} \cdot D _{2,tr} + 0.28] \sigma _{x, b,0}, &
         $\sigma _{x,0} \leq A _{T _2} < l _c$; \cr
         [{k_2 - 1\over k _2} {\sigma _{x,0} \over \sigma
         _{x,b,0}} + \sqrt {k _2\over k_2 - 1} - \pi (k _{2} -1)
         \psi (k _2, \xi _{2,c}) + 0.28] \sigma _{x, b,0}, &
         $A _{T _2} > l _c, \sigma _{x,0}$, \cr }
         \end{equation} 
where $A _{T _2}$ is the amplitude of displacement of the second
target; $\xi _{2,st}$, the parameter $\xi _{2}$ at the moment of the
second target stop; $D _{2,tr} = A _f/A _0$, the coefficient of the
decompression of the amplitudes of betatron oscillations of particles
at the second stage of cooling; $l _c = \pi k _2 \psi (k _2, \xi
_{2,c}) A _0$, the way of movement of the second target, for which the
instantaneous orbits of particles having maximum initial amplitudes of
betatron oscillations will be deepened into the target on the depth
higher then their final amplitudes of betatron oscillations; the symbol
$\leq$ is because of we have got the maximum value $0.28$ in (3.5) for
the value $\Delta \sigma _{x,\varepsilon, 1}$.

Notice that the Eq. (3.2) does not take into account that the target
pass of a finite distance per one turn $\delta x _{T _2} = |v_{T
_2}|\cdot T$, where $T = 1/f$ is the period of the particle revolution
around its orbit in the storage ring. When

                \begin{equation} \delta x_{T _2}> \sigma _{x, b,0},
                \end{equation}    
$|v _{T _2}| \geq |\dot x _{\eta \, in}|$ then all instantaneous orbits
of the particle beam can enter the target at the distance $x _{\eta } -
x _T > \sigma _{x, b,0}$, that is, all at once under conditions
$\partial A/\partial t = 0$ ($\varphi _2 = \pi$). In this case there
will not be any heating process in the transverse plane. This case can
be realized easier if we do a high-degree cooling of the particle beam
for the first stage of cooling, and when the target is installed at the
straight section with low $\beta $ - function and high dispersion
function at the second stage of cooling.

We considered the example of the transverse heating and longitudinal
cooling at the second stage for illustration. Detailed numerical
calculations of the amplitude of betatron oscillations increase and
damping of the energy spread of the beam at the second stage of cooling
are presented in the Appendix B.

The energy spread of the particle beam being cooled in the storage ring

        \begin{equation}     
        {\sigma _{\varepsilon } \over \varepsilon } > max\left(
        {\varepsilon _{loss} \over Mc^2 \gamma},  \hskip 5mm{\delta r
        \over \overline R}, \hskip 5mm ({\Delta \omega \over
        \omega})_{nat}\right), \end{equation}
where $\delta r$ is the length of the slope of the target edge;
$\overline R = cT$ the average radius of the ring; $\varepsilon
_{loss}$, the average loss of the particle energy per one event of
particle-photon interaction. The value ${\delta r /\overline R}$
influence on the energy spread mainly at the second stage of cooling.

The damping times of the particle beam at the first and second stages
of cooling in the transverse and longitudinal planes are

        \begin{equation}     
        \tau _{x} = {\sigma _{eq}\over k_{1} P},
        \hskip 5mm
        \tau _{s} = {\sigma _{\varepsilon \, in} \over P},
        \end{equation} 
where in the smooth approximation $\sigma _{eq} = \varepsilon \sigma
_{x,b, 0}/\alpha \overline R$ the energy interval corresponding to
the energy spread of the particle beam whose instantaneous orbits are
distributed through the interval of radii $\sigma _{x, b,0}$;
$k_{1} \sim 0.1 \div 0.2$; $\alpha $, the momentum compaction function,
$P$ is the power of the particle energy loss.

The transverse emittances of beams are proportional to theirs damping
times. It means that the emittance of a beam in the plane "i" in the
two-dimensional method of cooling $\epsilon _i ^{(2)}$
is equal to the emittance corresponding to a three-dimensional one
$\epsilon _i ^{(3)}$ multiplied by the ratio of their damping times

        \begin{equation}
        \epsilon _i ^{(2)} = \epsilon _i ^{(3)}{\tau _i ^{(2)}
        \over \tau _i ^{(3)}}.  \end{equation} 

The described process of transverse cooling is based on particle
interactions with external and internal targets. Similar interactions
were described in 1956 by O'Neil \cite{oneil}. However, the targets in
that case were motionless and could not lead to cooling\footnote{
Internal target could be rotated out of the medium plane only to
prevent the particle beam losses.}. They could be used for injection
and capture of only one portion of particles. For the purpose of the
multi-cycle injection and storage of heavy particles O'Neil, in
addition to targets, suggested the ordinary three-dimensional
ionization cooling based on a thin hydrogen target jet situated in the
working region of the storage ring.

       \begin{center} \bf Discussion \end{center}

The dynamics of positions of instantaneous orbits and amplitudes of
betatron oscillations of particles interacting with a target strongly
depends on the target velocity when the instantaneous orbits are
deepened into the target on the depth less then the amplitude of
particle betatron oscillations (see (3.2)). Moreover, the moving target
begin interact with particles of the beam located at different
instantaneous orbits at different moments of time and that is why can
compress or decompress the spread of these orbits. These features of
interaction of moving target can be used for improving existing and
adopting new schemes of enhanced three-dimensional cooling.

We have found (see (B.6), (3.4)), that at the first stage of cooling
there is the significant decrease of amplitudes of betatron oscillations
(transverse cooling) and, at the same time, a greater increase of the
spread of instantaneous orbits (longitudinal heating). If the degree of
transverse cooling (B.6) is defined by the coefficient of compression
$C _{1,tr} = A _0/A _f = \sqrt{(1 + |k _1|)/|k _1|}$ then, according to
(3.4), the increase of the spread of the instantaneous orbits of the
beam (decompression) will be $D _{1,l} \simeq C _{1,tr} ^{2}$ times.

At the second stage of cooling there is a significant decrease of the
spread of instantaneous orbits of particles defined by the compression
coefficient $C _{2,l} = \sigma _{x,\varepsilon,0}/ \sigma _{x,
\varepsilon ,f}$ and, at the same time, lesser value of increase of
amplitudes of betatron oscillations (see (3.5), (B.8), Tables 8, 9).
If the condition (3.6) is not fulfilled then the degree of the
transverse heating (decompression) can be about the square root of the
degree of the longitudinal cooling ($D _{2, tr} \simeq \sqrt{C _{2,
l}}$), when $\xi _{2, st} \simeq 0$ (see Table 8,9) or when $\xi _{2,
st} = -1$, $A _{T _2} \simeq \sigma _{x, \varepsilon, 0} \gg \sigma
_{x, b,0}$ (see Table 9 at $\xi _{2, st} = -1, A _{T _2} = 101 \sigma
_{x.b.0}$). In the last case the degree of selectivity of interaction
of a target with instantaneous orbits is greater. It can be realized
easier if we will locate the target in the straight section of the
storage ring with a low-beta and high dispersion function. When the
condition (3.6) is fulfilled then heating process can be neglected at
all.

The successive application of two stages of the two-dimensional cooling
in tern will lead to cooling of the particle beam in both degrees of
freedom only in the case when the condition (3.6) is fulfilled.

When the interaction of the particle beam and the target has the
resonance or threshold nature then we can use transverse cooling of
particles at the first stage of the two dimensional method of cooling
and then to use one dimensional method of longitudinal cooling at the
second one (see section (3.2.1)).

At the second stage, contrary to the first one, the degree of
longitudinal cooling can be much greater then the degree of heating
in the transverse plane. That is why we can use the emittance
exchange between longitudinal and transverse phase spaces (say
using the synchro-betatron resonance) and such a way to have
enhanced two-dimensional cooling of the particle beam based on
the second stage of cooling only. In this case the first stage
of cooling can be omitted and the second one can be repeated
many times (see sections (3.2.2), (3.2.3)).

\subsubsection{Enhanced laser cooling of ion beams in storage rings}

In the two-dimensional method of ion cooling based on the selective
resonant interaction, as the targets one can use two different laser
beams. The first broadband laser beam must have sharp geometrical
internal boundary. Metal screens can be used at the exit of the laser
beams from an optical resonators to produce the extracted laser beams
with sharp edges\footnote{ Production of laser beams with sharp edges
in resonators is another problem which would be solved.}.  The second
laser beam must overlap the ion beam as a whole, have sharp frequency
edge or can be monochromatic one and must have scanning central
frequency. In this case the ordinary one-dimensional resonance laser
cooling will take place at the second stage (see Section 3.1). At that
we can start from the second stage.  Then the first stage and the
second one must be repeated.

{\it Example 1.} The two-dimensional cooling of a hydrogen-like beam of
$^{207}_{82}Pb^{+81}$ in the CERN LHC through the backward Rayleigh
scattering of photons of two laser beam targets. The broadband laser
beams overlap the ion beam, have sharp frequency edges, and scanning
central frequencies.

The relevant parameters of LHC and the beam in LHC are: $2\pi \overline
R = 27$ km, $f = 1.11 \cdot 10 ^{4}$ Hz, $\alpha = 2.94 \cdot 10^{-4}$,
$<\beta_x>\, = 0.5 m$, $\gamma = 3000$, $Mc^2\gamma = 575$ TeV, $\sigma
_{\varepsilon ,0}/\varepsilon = 2\cdot 10^{-4}$ ($\sigma _{\varepsilon
,0} = 1.15\cdot 10^{11}$ eV), the value $\sigma _{x,b,,0} = \sigma _{x,
\varepsilon ,0} = 1.2\cdot 10^{-2}$ cm.

The relevant characteristics of the hydrogen-like ($f_{12} = 0.416$,
$g_1$ = 1, $g_2$ = 3) lead ions are:  the transition between the $1S$
ground state and the $2P$ excited state of the particle corresponds to
the value of the resonant transition energy $\hbar \omega _{tr} = 68.7$
keV, $\lambda _{tr} = 1.8 \cdot 10^{-9}$ cm, $(\Delta \omega/ \omega)
_{nat} = 2.72 \cdot 10^{-4}$.

The relevant parameters of a laser: the laser wavelength $\lambda _L =
4\pi c\gamma /\omega _{tr} = 1080 \AA$, $\hbar \omega _L = 11.49$ eV,
the bandwidth of the laser beam $(\Delta \omega /\omega )_L = 5\cdot 10
^{-4}$, the r.m.s. transverse laser beam size at its waist $\sigma _L =
1.52 \cdot 10^{-2}$ cm, the Rayleigh length $z_R = \pi \sigma
_L^2/\lambda _L = 67.2$ cm, $l_{int} = 2z_{R} = 135$ cm, the power of
the laser beam is $P_L = 400$ W.

Under this condition the average energy of the scattered photons
$<\hbar \omega ^s> = \varepsilon _{loss} = \gamma \hbar \omega _{tr} =
206$ MeV, $I_{sat} = 4.94 \cdot 10^{12}$ W/cm$^2$,  $I_L = 5.5 \cdot
10^{5}$ W/cm$^2$, $D = 1.1 \cdot 10^{-7} \ll 1$, $\overline \sigma =
1.32 \cdot 10^{-18}$ cm$^2$, $\Delta N_{int} = 3.52 \cdot 10 ^{-3}$, $P
= 8.04 \cdot 10^{9}$ eV/s.  According to (3.1), (3.7) (3.9) the
longitudinal damping time $\tau _s \simeq 14.3$ sec., the transverse
damping time $\tau _{x}| _{k_{1} = 0.1} \simeq 143$ sec., the total
damping time $\tau = 157.3$ sec., the limiting relative energy spread
$\sigma _{\varepsilon }/\varepsilon \simeq (\Delta \omega/ \omega)
_{nat} \simeq \sigma _{\varepsilon, \,0 }/\varepsilon $, $l _{dec}
\simeq 4.2 \cdot 10 ^{-3}$ cm. According to (2.1), (2.3), the damping
time and the emittance of the ion beam in the case of three-dimensional
ion cooling method are: $\tau _x = 1.43 \cdot 10 ^5 sec.$, $\varepsilon
_x = 3\cdot 10 ^{-15}$ m$\cdot$rad\footnote {The damping time of bunched
ion beams can be done less at the same average power $P$ \cite {prl}.}.
It means that in the two-dimensional method of ion cooling, according
to (3.9), $\varepsilon _x = 2.86 \cdot 10 ^{-16}$ m$\cdot$rad.

Notice, that at the first stage of the resonance ion cooling the width
of the frequency band can be small. It can correspond to the spread of
the instantaneous orbits $\Delta x _{\eta} \simeq A _f$. In this case
the amplitude of betatron oscillations of particles will be in time to
be cooled before their instantaneous orbits will go out of the spread
and will be stopped without interaction with the laser target. Heating
of the particle beam in the transverse plane at the another side of the
laser target will be absent.

The two-dimensional method of laser cooling of ion beams will work in
the case when the RF system of the storage ring is switched on as well.
In this case the scanning of the frequency of the accelerating fields
can be used instead of moving targets. At that the laser targets $T _1$
and $T _2$ have to be switched on and off in turn.

\subsubsection {Enhanced laser cooling of electron and proton beams}

To produce the enhanced laser cooling of electron and proton beams
based on the backward Compton scattering we are forced to use laser
beams with sharp geometrical boundaries.
We can not neglect synchrotron radiation of electron beams in the
guiding magnetic fields of lattices of storage rings. That is why we
are forced to coll such beams in the radio frequency buckets.  Cooling
of the particle beams in buckets is another problem which can be
considered in a separate paper. Obviously this problem can be solved
when only second stage of cooling is used under conditions of
synchro-betatron resonance.

Here we would like to notice that when the spread of amplitudes of
betatron oscillations is near to zero then we can take the parameter
$k_{2} = 1$. In this case the cooled beam according to (3.5) will be
monochromatic. It means that we can add the second stage of cooling
to the case of the three dimensional laser cooling of electron beams
considered in \cite{zhirong} one time and then extract the beam with
low both transverse and longitudinal emittances.

\subsubsection{Enhanced ionization cooling of muon beams in storage
rings}

Muons have rather small lifetime ($ \sim 2.2 \mu sec$) in their rest
frame. They can do about $3\cdot 10^3$ turns only in the strong
magnetic field ($\sim 10$ T). That is why the muon beams require
enhanced cooling. Muons have no nuclear interactions with the material
medium of the target.  That is why they have no problems with the
inelastic scattering in the target.

The enhanced method of the two-dimensional cooling of particle beams in
storage rings can be used in the case of muon cooling as well. Two
material medium targets are to be placed in the inner and external
sides of the working region of the storage ring. The use of thick
targets can lead to high rate of the ionization energy losses of muons
in the targets. The kick/bump can be used for the displacement of the
instantaneous orbits forth and back with high velocity ($\sim \dot
x_{\eta \, in}$) instead of the displacement of targets.

If the velocity of movement of the instantaneous orbit $\dot x_{T _2}$
by a kick in the direction of the external target is slightly greater
then the velocity of movement of the orbit in the target ($k _2 = \dot
x_{T _2}/\dot x_{\eta \, in} \simeq 1$) then during the second stage of
longitudinal cooling the instantaneous orbits will be gathered at the
inner orbit position in a damping time (3.1), (3.8), where $P$ is the
average power of the ionization energy loss. At this moment the kick
must be switched off for a short time. In this case, the muon beams
will be cooled down to high degree in the longitudinal plane. To avoid
the heating of the beam in the transverse plane we must work under
condition (3.6), that is $\dot  x_{T _2} \cdot T > \sigma _{x, \,b, \,
0}$\footnote{We suppose that after acceleration the transverse and
longitudinal emittances of muon beams are much less then the
corresponding acceptances of the storage ring.}.

The enhanced transverse muon cooling will take place when the
instantaneous orbit will be moved to the muon beam with the velocity
$\dot x_{T _1} \simeq 0.1 \cdot x_{\eta \, in}$ (see section 3.1).

The two-stage cooling process can be repeated several times to increase
the degree of cooling. To keep the position of instantaneous orbits of
particles at central part of the working region of the storage ring
after cooling stages the magnetic field of the storage ring can be
decreased in time.

Another version of the two-dimensional muon cooling is based on using
of the second stage of cooling only and the coupling of the
longitudinal and transverse planes through the synchro-betatron
resonance. In this case the first stage of cooling can be omitted.

\subsubsection {Enhanced cooling of ion beams through the $e^{\pm}$ pair
production}

The three-dimensional method of laser cooling of ion beams based on the
nonselective interaction of counterpropagating ion and photon beams
through the $e^{\pm}$ pair production were considered in \cite {bw}.
More effective two-dimensional method of cooling based on the threshold
phenomena of the $e^{\pm}$ pair production in Coulomb fields of ions
was considered in \cite{bkw}. Below we will consider the latter case.

The cross-section of the electron-positron pair production has
the form

     $$\sigma |_{\hbar \omega - 2m_e c^2 \leq m_e c^2} \sim {\pi \over
   12} Z^2 \alpha r_e ^2 \left({\hbar \omega -2 m_e c^2 \over m_e
   c^2}\right)^3,$$

      \begin{equation}
      \sigma |_{\hbar \omega \gg 2m_e c^2} \simeq {28 \over 9} Z^2
      \alpha r_ e ^2 [\ln {2\hbar \omega \over m_e c^2} - {109\over 42}
      - f(\alpha Z)],
      \end{equation} 
where $Z$ is the atomic number of the ion, $f(\nu)/\nu ^2 \simeq 1.203
- \nu ^2$ \cite{blp}.

The cross-section (3.10) has a threshold photon energy $\hbar \omega
_{thr}^{'} = 2m_e c^2$ in the ion rest frame or $\hbar \omega _{thr} =
2m_e c^2/(1 + \beta _{thr})\gamma _{thr}$ in the laboratory reference
frame, where the ion threshold relativistic factor $\gamma _{thr} = [1
+ (\omega _{thr}^{'}/\omega _L)^2] /2(\omega _{thr}^{'}/ \omega _L)$.

Below we will consider an example of the two-dimensional laser cooling
of the lead ion beam through the $e^{\pm}$ pair production.

{\it Example 2.} The two-dimensional cooling of the fully stripped ion
beam of lead $^{207}_{82}Pb^{+82}$ in the CERN LHC through the
$e^{\pm}$ pair production.

The relevant parameters of LHC and the beam in LHC are the same
as in the previous example:  $2\pi \overline R = 27$ km, $f =
1.11 \cdot 10 ^{4}$ Hz, $\alpha = 2.94 \cdot 10^{-4}$, $\gamma =
3000$, $Mc^2\gamma = 575$ TeV, $\sigma _{\varepsilon
\,in}/\varepsilon = 2\cdot 10^{-4}$ ($\sigma _{\varepsilon \,in}
= 1.15\cdot 10^{11}$ eV), the value $\sigma _{\varepsilon b} =
\sigma _{\varepsilon \,in}$, $\sigma _{x \, b} = \sigma _{x
\,\varepsilon} = 1.2\cdot 10^{-2}$ cm.

In this method of cooling $\hbar \omega _{thr}^{'} = 2m_e c^2 = 1.02
MeV$, $\lambda _{thr} ^{'} = 0.01217 \AA$, $\hbar \omega _{thr} = 2m_e
c^2/(1 + \beta)\gamma = 170$ eV, $\lambda _{thr} = 73 \AA$.

The relevant parameters of a laser: wavelength $\lambda _L =
36.5 \AA$, $\hbar \omega _L = 340$ eV, the bandwidth $(\Delta \omega
/\omega )_L < 10 ^{-4}$, the r.m.s. transverse size at a beam waist
$\sigma _L = 1.52 \cdot 10^{-2}$ cm, the Rayleigh length $z_R = \pi
\sigma _L^2/\lambda _L = 1989$ cm, $l_{int} = 2z_{R} = 3978$ cm; the
power $P_L = 400$ W.

In this case the average energy loss per one event of pair production
$\varepsilon _{loss} \simeq 2\gamma m_e c^2 = 3.06$ GeV, $I_L = 5.48
\cdot 10^{5}$ W/cm$^2$, $\sigma = 8.1 \cdot 10^{-24}$ cm$^2$, $\Delta
N_{int} = 2.69 \cdot 10 ^{-9}$, $P = 6.43 \cdot 10^{10}$ eV/s., the
longitudinal damping time $\tau _s \simeq 1.58 \cdot 10^5$ s., the
damping time for the betatron oscillations $\sim$ 10 times greater.
According to (3.7) the limiting relative energy spread is $\sigma
_{\varepsilon}/\varepsilon \simeq 5.3 \cdot 10^{-6}$.

The average power of the X-ray laser in the considered example is
rather high but it can be realized in future FELs \cite{srn}. The
X-ray FELs are proposed to operate in the pulsed regime. In this case
the damping time or/and the power of the laser can be decreased
essentially (two-four orders) if we use ion beam gathered in short
bunches separated by a long distances (duty cycle $\sim 10^2 \div 10
^4$), and use interaction regions with smaller diameters of ion
and photon beams.

\section{Cooling of electron and ion beams in linear accele\-rators}

Electron beams can be cooled in linear accelerators if external
fields (undulators, electromagnetic waves) producing radiation
friction forces will be distributed along the axes of these
accelerators. The physics of cooling of electron beams under
conditions of linear acceleration is similar to one in storage rings
where external fields are created in the dispersion-free straight
sections of the rings, and the synchrotron radiation in bending magnets
of the rings can be neglected (see Section 2). The element of the
irreversibility takes place in this case too. The electron momentum
losses are parallel to the particle velocity and therefore include
transverse and longitudinal momentum losses. The reacceleration of the
electron beams in accelerating structures of linear accelerators
restores the longitudinal momentum.  The transverse emittance is
reduced by $1/e$ with as little as $2\varepsilon $ of the total energy
exchange. The expected emittances of cooled beams are small in both
transverse directions.

First, the effect of undulator/wiggler radiation damping on the
transverse beam emittance was studied by A.Ting and P.Sprangle for
linear accelerators based on inverse free-electron lasers \cite{ting}.
In \cite{dikan}, the same effect applied to the case of the radio
frequency linear accelerators is considered.  In \cite{spr} and later
in \cite{telnov}, the case of linear acceleration was investigated,
where a laser beam was used instead of a wiggler.  General formulas in
this cases are similar. Some peculiarities are in the hardness of the
emitted radiation which determines the energy spread and transverse
emittance of being cooled beams. The hardness of the backward scattered
laser radiation is more high then undulator/wiggler radiation. That is
why laser beams for damping can be used at small ($\sim 10^3 \div 10^4
$ MeV) electron energies.  A strong focusing of electron and laser
beams at the interaction point is necessary in this case. Damping
wigglers can be used at high energies ($10 \div 100$ GeV) and in the
limits of more long distances along the axis of the accelerator (about
some kilometers).

The analogies with the enhanced particle cooling in linear accelerators
are possible as well when bending magnets will be used for dispersion
separation of particles and selective cooling.  Monochromatization of
ion beams can be realized by broadband lasers with sharp frequency edge
located at the exit of the linear accelerator.

                        \section{Conclusion}

The fundamental ideas of cooling of particle beams based on a friction
were invented during the past five decades. First the synchrotron
radiation damping/cooling was developed theoretically. The theory of
the synchrotron radiation damping is used during designing of the
$e^{\pm}$ synchrotrons and storage rings for the $e^{\pm}$ circular
colliders and next generations of the synchrotron and undulator
radiation sources.  Activity in this field have led to the development
of storage ring lattices (straight sections with high- and low-beta
functions and zero momentum compaction factors, using of magnets with
low magnetic fields and large bending radii).  The necessity in
increasing of luminosity of the $p, \overline p$ colliders have led to
a development of electron and stochastic methods of cooling of heavy
particles. The development of the idea of the inertial confinement
fusion have led to a necessity of ion cooling of non-fully stripped ion
beams through the intrabeam scattering \cite[a]{ rubbia}.  The
development of $\mu$ colliders stimulated development of schemes based
on well-known methods of ionization friction. Laser cooling in gaps was
naturally extended to the enhanced one-dimensional laser cooling in
storage rings, to the enhanced three-dimensional cooling through a
synchro-betatron resonance or through the longitudinal-transverse
dispersion in storage rings.

In this review we have presented different methods of cooling of
relativistic particle beams in a single particle approximation. We
hope that the development and adoption of these methods will lead to
the next generation of storage rings for colliders of different
particles, new light sources in optical to X-ray and $\gamma$ -ray
regions \cite{pac}, \cite{zhirong}, ion fusion \cite[a]{rubbia},
sources of gravitational radiation in IR and more hard regions
\cite{bes2}, and so on.

\newpage
\addcontentsline {toc} {section} {\protect\numberline
    {6 \hskip 2mm References}}

\newpage
\addcontentsline {toc} {section} {\protect\numberline
{7 \hskip 2mm Appendix A}}
\begin{center}
\renewcommand{\appendixname}{\large \bf Appendix A}\appendixname
\end{center}
\setcounter{section}{0}
\setcounter{equation}{0}
\def\theequation{A.\arabic{equation}}

In a smooth approximation, the movement of a particle relative to its
instantaneous orbit position $x_{\eta}$ is described by

       \begin{equation}             
       x^{'} = A\cos(\Omega t+\varphi).
       \end{equation}
where $x^{'} = x - x_{\eta}$.

The amplitude of betatron oscillations of the particle $A _0 = \sqrt
{x^{'\, 2}_0 + \dot x _0^{'\, 2}/\Omega ^2}$, where $x^{'}_0$ is the
particle deviation from the instantaneous orbit at the moment $t_0$ of
change of the particle energy in a target (laser beam, material
medium), $\dot x^{'}_0 = - A \Omega \sin (\Omega t + \varphi )$ the
transverse velocity of the particle. After the interaction, the
position of the instantaneous orbit will be changed at a value $\delta
x_{\eta}$. The particle will continue its movement relative to a new
orbit. Its deviation relative to the new orbit will be $x^{'}_0 -
\delta x _{\eta}$, and the angle will not be changed. The new amplitude
will be $A_1 =\sqrt {(x^{'}_0 - \delta x _{\eta})^2 + \dot x^{'\,
2}_0/\Omega ^2}$.  The change of the square of the amplitude

       \begin{equation}
       \delta (A)^2 = A_1^2 - A_0^2 = - 2x^{'}_0\delta x _{\eta} + (\delta
       x _{\eta}) ^2.
       \end{equation}          

When $|\delta x _{\eta}| \ll |x^{'} _0| < A_0$, the second term in
(A.2) can be neglected\footnote{In the framework of classical
electrodynamics particles loose theirs energy continuously. That is why
the value $(\delta x_{\eta})^2$ in (A.2) in this framework can be
neglected.  The averaged value $<2P(x^{'})x^{'}> \ne 0$ when the
gradient of the energy loss $\partial P(x^{'})/\partial x^{'} \ne 0$
(see Appendix B).  Both damping and antidamping of betatron
oscillations of particles in storage rings can be effective in this
case \cite{kolom-leb}. The term $<P(x^{'}) (\delta x _{\eta}) ^2>$
leads to the excitation of betatron oscillations only in the case of a
intermittent random energy loss.}. In this case the value $\delta (A)^2
\simeq - 2x^{'}_0\delta x _{\eta}$ and $\delta A = - (x _0 ^{'}/A)
\delta x _{\eta}$.  The amplitude of betatron oscillations of the
particle will be changed by the low $|\delta A| \simeq |\delta x
_{\eta}|$ when $|x^{'}_0| \simeq A$.  It means that the particle will
change its amplitude of oscillations proportional to the number of
passages $N$ of the particle through the target when the instantaneous
orbit of the particle is at a distance of about their amplitude ($\sim
A$) away from the target. This is the highest rate of increase in the
amplitude of the particle oscillations.

When the target is located at the external side of the working region
of the storage ring $x_{T _2} > 0$, the instantaneous orbit position
$x_{\eta} < x_{T _2}$, the particle enter the target under conditions
of deviations $x^{'}_0 > 0$, and when the energy loss of the particle
leads to the decrease of its instantaneous orbit position ($\partial x
_{\eta}/\partial \varepsilon > 0$) then the amplitude of radial
betatron oscillations of the particle will be increased (heating
conditions). In the opposite case when the target is located at the
inner side of the working region of the storage ring $x_{T _1} < 0$,
$x_{\eta} > x_{T _1}$, and $x^{'}_0 < 0$ the amplitude of radial
betatron oscillations of the particle will be decreased (cooling
conditions).

The term $-2x _0^{'}\delta x_{\eta}$ in (A.2) determines the classical
damping (antidamping) processes in particle beams of storage rings.
The value $\delta x _{\eta} = D_x \Delta p/p$, where $D_x$ is the local
dispersion function \cite{kolom-leb}, $p = Mc\beta \gamma$ the momentum
of the particle. This means that the scheme works when the dispersion
function $D _x \ne 0$. The greater $D_x $ the greater the rate of
cooling.

\newpage
\addcontentsline {toc} {section} {\protect\numberline
{8 \hskip 2mm Appendix B}}
\begin{center}
\renewcommand{\appendixname}{\large \bf Appendix B}\appendixname
\end{center}
\def\theequation{B.\arabic{equation}}
\setcounter{equation}{0}

From the definition of $\xi _{1, \,2}$ we have a relation $x _{\eta} =
x _{T _{1,\,2}} - \xi _{1, \,2} A(\xi _{1, \,2})$. The time derivative
$\partial x _{\eta}/\partial t = v _{T _{1,2}} - [A + \xi _{1,\,2}
(\partial A/ \partial \xi _{1,\,2})] \partial \xi _{1,\,2} /\partial
t$, where $v _{T _{1,2}} = dx _{T _{1,2}}/dt$ is the velocity of the
target. Equating this value to the second term in (3.2) we will receive
the time derivative

        \begin{equation} 
        {\partial \xi _{1,\,2}\over \partial t} = {\dot x _{\eta \,in}
        \over \pi} {\pi k _{1,2} - \Delta \varphi _{1,\, 2} \over A(\xi
        _{1,\,2}) + \xi _{1,\,2} (\partial A / \partial \xi
        _{1,\,2})}.  \end{equation}
Using this equation we can transform the first value in (3.2) to the
form

     $$\pm sinc \Delta \varphi _{1,2} (\xi _{1,\,2}) = {\partial A\over
     \partial \xi _{1,\, 2}}{\partial \xi _{1,\,2}\over \partial t}/
     {\partial x _{\eta}\over \partial t} = {\pi k _{1,2} - \Delta
     \varphi _{1,2} \over [A + \xi _{1,\,2}(\partial A/\partial \xi
     _{1,\,2})] \Delta \varphi _{1,2}} {\partial A /\partial \xi
     _{1,\,2}}.$$ which can be transformed to

     $${\partial \ln A \over \partial \xi _{1,\,2}} = {\pm \sin
     \Delta \varphi _{1,2} \over \pi k _{1,2} - (\Delta \varphi
     _{1,2} \pm \xi _{1,\,2} \sin \Delta \varphi _{1,2})}.$$

When velocities of targets $v _{T _{1\, 2}}$ and velocities of
instantaneous orbits in the targets $\dot x _{\eta \, in}$  are
constant (targets have constant thickness) then the received equation
leads to the law of change of the amplitudes of betatron oscillations
of particles in the storage ring

       \begin{equation}
       A _f = A(\xi _{1,\,2,\,f}) = A _0 \exp \int _{\xi _{1,2,0}} ^
       {\xi _{1,2,f}}{\pm \sin \Delta \varphi _{1,2} d\xi _{1,\,2,\,f}
       \over \pi k _{1,2} - (\Delta \varphi _{1,2}\pm \xi _{1,\,2}
       \sin \Delta \varphi _{1,2})},
       \end{equation} 
where the labels $0$, $f$ correspond to the initial time $t_0$ and
observation time $t _f$ accordingly.

The time dependence of the amplitudes and positions of the
instantaneous orbits of particles are determined by (3.2) through the
parameter $\xi (t)$. This parameter is determined by (B.1) and (B.2).
Substituting the values $A$ and $\partial A/ \partial \xi _{1,\, 2,
\,f}$, which are determined by (B.2), in (B.1) we can find the
connection between time of observation and parameter $\xi _{1,\,2}$

       \begin{equation}
       t - t _0 = {\pi A _0 \over |\dot x _{\eta \,in}|} \psi
       (k _{1,2}, \xi _{1,\,2, \,f}),   \end{equation} 
where

$$\psi (k _{1,2}, \xi _{1,\,2,\,f}) = \int ^{\xi _{1,\,2, \,f}} _{\xi
_0}{ -[A(\xi _{1,\,2, \,f})/ A_0] d\xi _{1,\,2, \,f} \over \pi k _{1,
\,2} - (\Delta \varphi _{1,\,2} \pm \xi _{1,\,2, \,f}\sin \Delta
\varphi _{1,\,2})}. $$

The equations (B.3) determine the time dependence of the functions
$\xi _{1,\,2}(t _f - t _0)$. The time dependence of the amplitudes
$A[\xi _{1,\,2} (t - t _0)]$ is determined by the equation (B.2)
through the functions $\xi _{1,\,2}(t - t_0)$ in a parametric form. The
time dependence of the position of the instantaneous orbit follow from
the definition of $\xi _{1,\,2}$

       \begin{equation}
       x_{\eta}(t - t _0) = x_{T _{1,\, 2}0} + v _{T_{1,\,2}}(t _f -
       t_0) - A[(\xi _{1, \,2} (t _f - t_0)]\cdot \xi (t _f - t _0).
       \end{equation} 

\newpage
        \begin{center} \bf First  stage of cooling \end{center}

The law of change of the amplitudes of particle betatron oscillations
is determined by the equation (B.2), which in the first stage of
cooling can be presented in the form

       \begin{equation}
       A _f = A _0 \exp \int _{\xi _0} ^{\xi _{1 ,\,f}}
       {- \sqrt{1 - \xi _1^2} d\xi _1 \over - \pi k _{1} + \pi -
       \arccos \xi _1 + \xi _1 \sqrt {1 - \xi _1 ^2}}.
       \end{equation} 

The dependence of the ratio $A _f/A _0$ of the final amplitude of
betatron oscillations of particles $A _f$ to the initial one $A_0$ on
the relative velocity $k _1 < 0$ of the target $T_1$ is described by
the equation (B.7). The numerical calculations of this dependence are
presented at the Fig.2 and in the Table 2 for the case $\xi _0 = - 1$,
$\xi _f = 1$.

\begin{figure}[hbt]
\centerline{\leavevmode\epsfxsize=150mm\epsfbox[92 370 556
722]{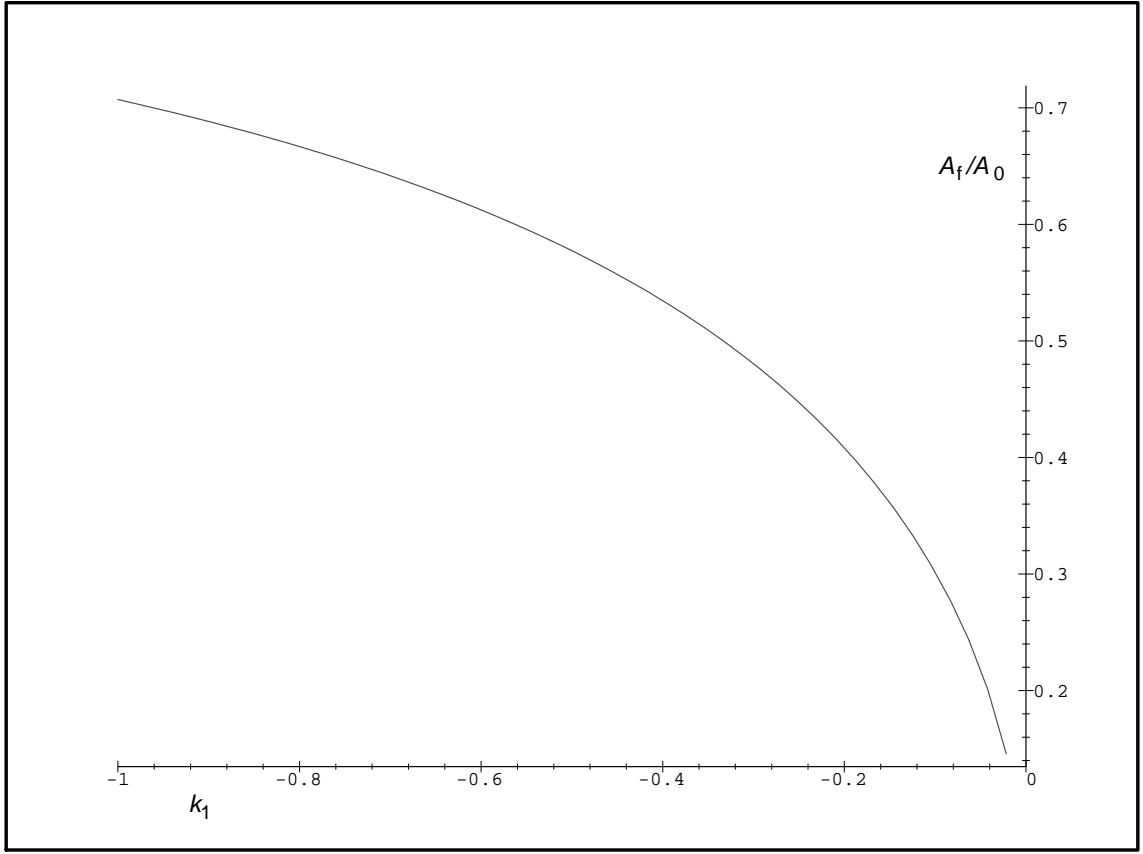}}
\caption{\small \it The dependence of the ratio $A _{f}/A_0$ on $k _1$.}
\vskip 4mm
Table 2
\vskip 5mm
\begin{tabular}{|l|l|l|l|l|l|l|}
\hline
$|k_1|$                   &0&0.2&0.4&0.6&0.8&1.0\\
\hline $C_1^{-1}=A_{f}/A_0$  &0&0.408&0.535&0.612&0.667&0.707\\
\hline $\sqrt{|k _1|/(|k _1|+1)}$ &0&0.408&0.535&0.612&0.667&0.707\\
\hline
\end{tabular}
\end{figure}

This dependence can be presented by the next approximate expression
(possibly this expression is the exact solution: precision of an
estimate $10 ^{-10}$).

       \begin{equation}
       A _{f} \simeq A _0 \sqrt{|k_1|\over |k _1| +1}.
        \end{equation} 

      \begin{center} \bf Second stage of cooling \end{center}

1) The law of change of the amplitudes of particle betatron oscillations
is determined by the equation (B.2), which in the second stage of
cooling can be presented in the form

       \begin{equation}
       A _f = A _0 \exp \int _{\xi _0} ^{\xi _{2,f}}{- \sqrt{1 - \xi
       _2^2} d\xi _2 \over \pi k _{2} - \arccos \xi _2+ \xi _2\sqrt {1
       - \xi _2^2}}.  \end{equation} 
where the parameter $\xi _{2,f}$ is determined by a moment $t_f = min
\{t _{st}, t _A\}$, where $t _{st} = t _0 + A _{T_2}/|v _{T_2}|$ is the
moment of the target $T _2$ stop, $ A _{T_2} /|v _{T_2}|$ is the
duration of movement of the target $T _2$ through its amplitude of
displacement $A _{T_2}$ and $t _A$, corresponds to the moment when the
instantaneous orbit of a particle will reach in the target the depth
equal its amplitude of betatron oscillations $A$, i.e. when $\xi _{2,A}
= \xi _{2,f}(t _A) = -1$. After this moment the amplitude of the
particle oscillations is not changed. We will consider here the cooling
of particle beams under conditions $\xi _0 = 1$. When $A _{T _2} < l
_{A}$ then, according to (B.7), the value $A _{f}$ has to be calculated
in the limits ($\xi _{2, {st}}, 1$), where $\xi _{2,{st}} > -1$
corresponds to the moment $t _{st}$.

The moment $t _A$ can be realized only when the relative velocity of
the second target $k _2 > 1$ and $t _{st} > t _A$. This moment,
according to (B.3), is determined by a moment $t _A = t _0 + \pi A _0
\psi (k _2, \xi _{2,c})/|x _{\eta, in}|$. During the time interval $t
_c - t _0$ the target $T _2$ will pass a way $l _{A _0} = |v _{T _2}|(t
_A - t _0) = \pi k_2 A _{0}$ $\psi (k _2, \xi _{2,c})$. The dependence
$\psi (k _2, \xi _{2,c})$ determined by (B.9) is presented in the Table
3. The value $l _A$ depends on initial amplitude of a particle betatron
oscillations and tend to maximum one $l _c = \pi k _2 \psi (k _2, \xi
_{2,c})\sigma _{x,b,0}$.

\begin{figure}[hbt]
\vskip 2mm
Table 3  \hskip 20 mm
\vskip 2mm
\begin{tabular}{|l|l|l|l|l|l|l|l|l|l|l|l|l|l|}
\hline
$k _2$ & 1.0 & 1.02 & 1.03 & 1.05 & 1.1&1.2 &1.3&1.4&1.5&1.7&2.0 \\
\hline
$\psi (k _2, \xi _{2,c})$
&$\infty$&13.80&9.90&6.52&3.71&2.10&1.51&1.18&0.980&
0.735&0.538\\
\hline
\end{tabular}
\end{figure}

\vskip 4mm

The ratio of a maximum amplitude of betatron oscillations of a particle to
the initial one $D _{2,tr,c} = A _{f,c} /A _0$ ($A _{f,c} = A _f(\xi
_{2,c}$) corresponding to the case $t _{st} > t _c$ on the
relative velocity $k_2$ of the second target is presented at the Fig.3
and in the Table 4.  According to calculations this ratio can be
presented by the next approximate expression (possibly this expression
is the exact solution:  precision of an estimate $10 ^{-10}$).

       \begin{equation}
       A _{f,c} \simeq  A _0\sqrt{k_2\over k_2 - 1}.
        \end{equation} 

2) The function $\psi (k _2,\xi _{2,f})$ according to (B.3) can be
presented in the form

       \begin{equation}
       \psi (k _2, \xi _{2,f}) = \int _{\xi _{2,f}} ^{1}{\exp
       \int ^{1} _{x} {[\sqrt{1 - t^2} / (\pi k _{2} - \arccos t +
       t\sqrt {1 - t ^2})]d\,t } \over \pi k _{2} - \arccos x + x \sqrt
       {1 - x ^2}}dx.  \end{equation} 

\begin{figure}[hbt]
\centerline{\leavevmode\epsfxsize=150mm\epsfbox[92 370 556
722]{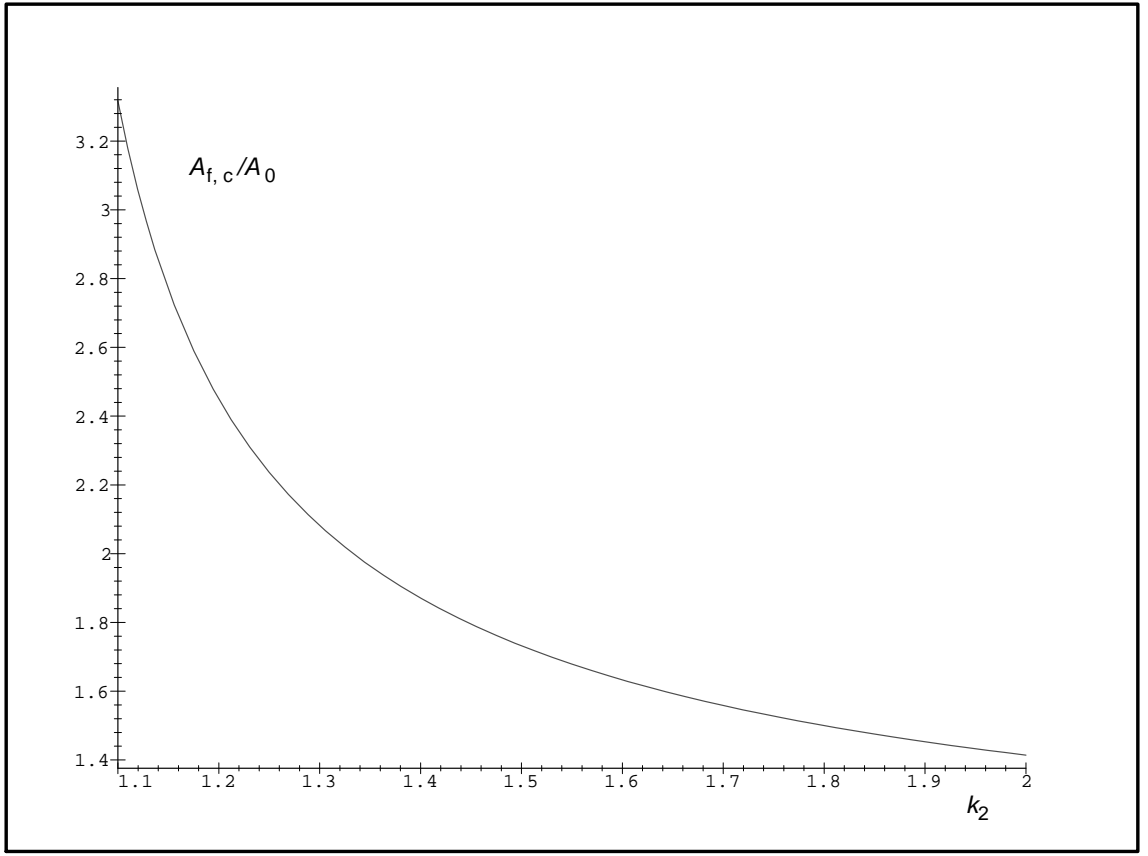}}
\caption{\small \it The dependence of the ratio $D _{2,tr,c} = A
_{f,c}/A_0$ on $k _2$.}

\vskip 4mm Table 4
\vskip 5mm \begin{tabular}{|l|l|l|l|l|l|l|}
\hline
$k_2$ & 1.0001 & 1.0010& 1.0100 & 1.1000 & 1.5000 & 2.0000 \\
\hline
$D _{2,tr,c} = A _{f,c}/A _0$ &$100.005$&31.64&10.04
&3.32&1.73&1.414 \\
\hline
$\sqrt {k _2/(k _2 - 1)}$ &$100.005$&31.64&10.04 &3.32&1.73&1.414 \\
\hline
\end{tabular}
\end{figure}

Numerical calculations of this function for the cases $k _2 = 1.0$, $k
_2 = 1.1$ and $k _2 = 1.5$ are presented in the Table 5, Table 6 and
Table 7 accordingly. It can be presented in the next approximate form

       \begin{equation}
       \psi (k _2, \xi _{2,f}) \simeq C _3(k _2)\psi ({1 - \xi _{2,f}
       \over k _2 + \xi _{2,f}}),  \end{equation} 
where $C _3(k _2) \simeq 0.492 - 0.680 (k _2 -1) + 0.484(k _2 - 1)^2 +
...$ , $\psi [({1 - \xi _{2,f}) / (k _2 + \xi _{2,f}})|_{k _2 = 1}
\simeq (1 - \xi _{2,f}) / (1 + \xi _{2,f}))$.

\begin{figure}[hbt]
\vskip -2mm
Table 5 \hskip 20 mm ($k_2 = 1.0$)
\vskip 2mm
\begin{tabular}{|l|l|l|l|l|l|l|l|l|l|l|}
\hline
$\xi _2$ & 1.0 & 0.5 & 0.2& 0&-0.2 & -0.5&-0.8&-0.9& -1.0 \\
\hline
$\psi(k _2,
\xi_2)$&$0$&0.182&0.341&0.492&0.716&1.393&4.388&10.187&-$ \infty$
\\ \hline \end{tabular} 

\vskip 2mm
Table 6 \hskip 20 mm ($k_2 = 1.1$)
\vskip 2mm
\begin{tabular}{|l|l|l|l|l|l|l|l|l|l|l|}
\hline
$\xi_2$ & 1.0 & 0.5 & 0.2&0 &-0.2 &-0.5& -0.8 & -0.9& -1.0 \\
\hline
$\psi (k _2, \xi_2)$&$0$&0.163&0.300&0.423&0.595 &1.033&
2.076&2.759 & 3.710 \\ \hline \end{tabular}
\vskip 4mm

\vskip 2mm
Table 7  \hskip 20 mm ($k_2 = 1.5$)
\vskip 2mm
\begin{tabular}{|l|l|l|l|l|l|l|l|l|l|l|l|l|l|}
\hline
$\xi _2$ & 1.0 & 0.5 & 0.2&0 & -0.2& -0.4&-0.6&-0.8&-1.0 \\
\hline
$\psi (k _2, \xi_2)$&$0$&0.116&0.202&0.273&0.359&0.466&0.602
&0.772 &0.980\\
\hline \end{tabular}
\end{figure}
\vskip 4mm

3) The evolution of instantaneous orbits and amplitudes of betatron
oscillations under the influence of the target pass the next moments.

a) First, the target $T _2$ will interact with particles having the
largest initial amplitudes of betatron oscillations $A _0 = \sigma
_{x,b,0}$ and the highest energies. The instantaneous orbit of these
particles, according to the definition of the function $\xi _2$ will be
changed by the low $ x _{\eta} ^{'} = x _{T_2} - \xi _{2,f} \sigma
_{x,b}(\xi _{2,f})$, where $x _{T_2} = x _{T_2,0} + v _{T_2}(t - t_0)$
up to the time $t = t _c$. At the same time instantaneous orbits $x
_{\eta} ^{"}$ of particles having the same maximum energy but zero
amplitudes of betatron oscillations will be at rest up to the moment $t
_0^{'} = t _0 + \sigma _{x,b,0}/|v _{T_2}|$. The orbit $x _{\eta} ^{'}$
at the moment $t _0^{'}$ will be displaced relative to the orbit $x
_{\eta} ^{"}$ by the distance $|\Delta x _{\eta}|$, where the difference
between the positions of orbits $\Delta x _{\eta}| _{t _0 \leq t \leq t
_0 ^{'}} = x _{\eta} ^{'} - x _{\eta} ^{"} < 0$. At the moment $t _0
^{'}$ the distance $|\Delta x _{\eta}|$ will be maximum. It will be
determined by the value

       \begin{equation}
       \Delta x _{\eta, 1} = \Delta x _{\eta}| _{t = t _0 ^{'}} =
       - \xi _2(t _0 ^{'})\cdot \sigma _{x,b}(t _0^{'}) < 0,
       \end{equation} 
where the parameter $\xi _2(t _0^{'})$, according to (B.3) and the
condition $|v _{T _2}|(t _0 ^{'} - t _0) = \sigma _{x,b,0}$, will be
determined by the equation $\psi [k _2, \xi _{2,f}(t _0 ^{'})] = 1/\pi
k_2$.

The value $\psi [k _2, \xi _{2,f}| _{k_2 \simeq 1} \simeq 1/\pi$, $\xi
_{2,f} (k _2, t _0 ^{'})|_{k _2 \simeq 1}$ $ \simeq 0.22$ (see Tables
5-7), $\sigma _{x,b}(t _0^{'}) = 1.26 \sigma _{x,b,0}$ and the distance
$|\Delta x _{\eta, 1}| \simeq 0.28 \sigma _{x,b,0}$. This distance will
be decreased with increasing $k _2$.

b) The instantaneous orbit of particles $x _{\eta} ^{"}$ inside the
time interval  $t _0^{'} <t \leq t _c$ will be changed by the low $x
_{\eta}^{"} = x _{T_2,0} - \sigma _{x,b,0} + \dot x _{\eta,\,in}(t -
t_0 - \sigma _{x,b,0}/|v _{T_2}|)$.  The difference between the
positions of two instantaneous orbits of particles having maximum and
zero initial amplitudes of betatron oscillations and equal initial
energies will be equal to $\Delta x _{\eta, 2} = [(k _2 -1)/k
_2][\sigma _{x,b,0} + v _{T _2}(t - t _0)] - \xi _{2,f} \sigma
_{x,b}(\xi _{2,{st}})$. At the moment $t _{st}$, i.e. at the position
of maximum displacement of the second target, when $t _{st} - t _0 = -
A _{T _2}/v _{T _2}$ the distance

       \begin{equation}    
       \Delta x _{\eta, 2}| _{t _0 \leq t _{st}\leq t _c} = -
       \left[{k _2 - 1\over k _2}\left({A _{T _2}\over \sigma _{x,b,0}}
       - 1\right) + \xi_{2,st} D _{2,tr}) \right]\sigma _{x,b,0},
       \end{equation}
where $D _{2,tr} = D _2(k _2, \xi _{2,{st}}) = \sigma _{x,b,st}/\sigma
_{x,b,0}$, $\sigma _{x,b,st} = \sigma _{x,b,f}| _{t = t _{st}}$, $A _{T
_2} = \pi k_2 \psi (k _2, \xi _{2,st}) \sigma _{x,b,0} \leq l _c$. The
typical dependence $D _{2,tr}$ defined by (B.7) is presented on the
Fig.4.

c) After the moment $t _c$ $(t _c < t < t _{st})$ the value (B.12) have
a maximum corresponding to $\xi _{2,st} = \xi _{2,c} = -1$, $A _{T _2}
= l _c$, $D _{2, tr, c} = \sqrt {k_2/(k _2 - 1)}$:

       \begin{equation}  
       \Delta x _{\eta,2}| _{t _c \leq t _{st}\leq \infty}  = \left[{k
       _2 - 1\over k _2} + \sqrt {k _2\over k _2 -1} -
       \pi (k _2 -1) \psi (k _2, \xi {2,c})\right]\sigma _{x,b,0}.
       \end{equation}

\vskip 10mm
\hskip 9mm
\begin{figure}[hbt]
\centerline{\leavevmode\epsfxsize=150mm\epsfbox[92 370 556
722]{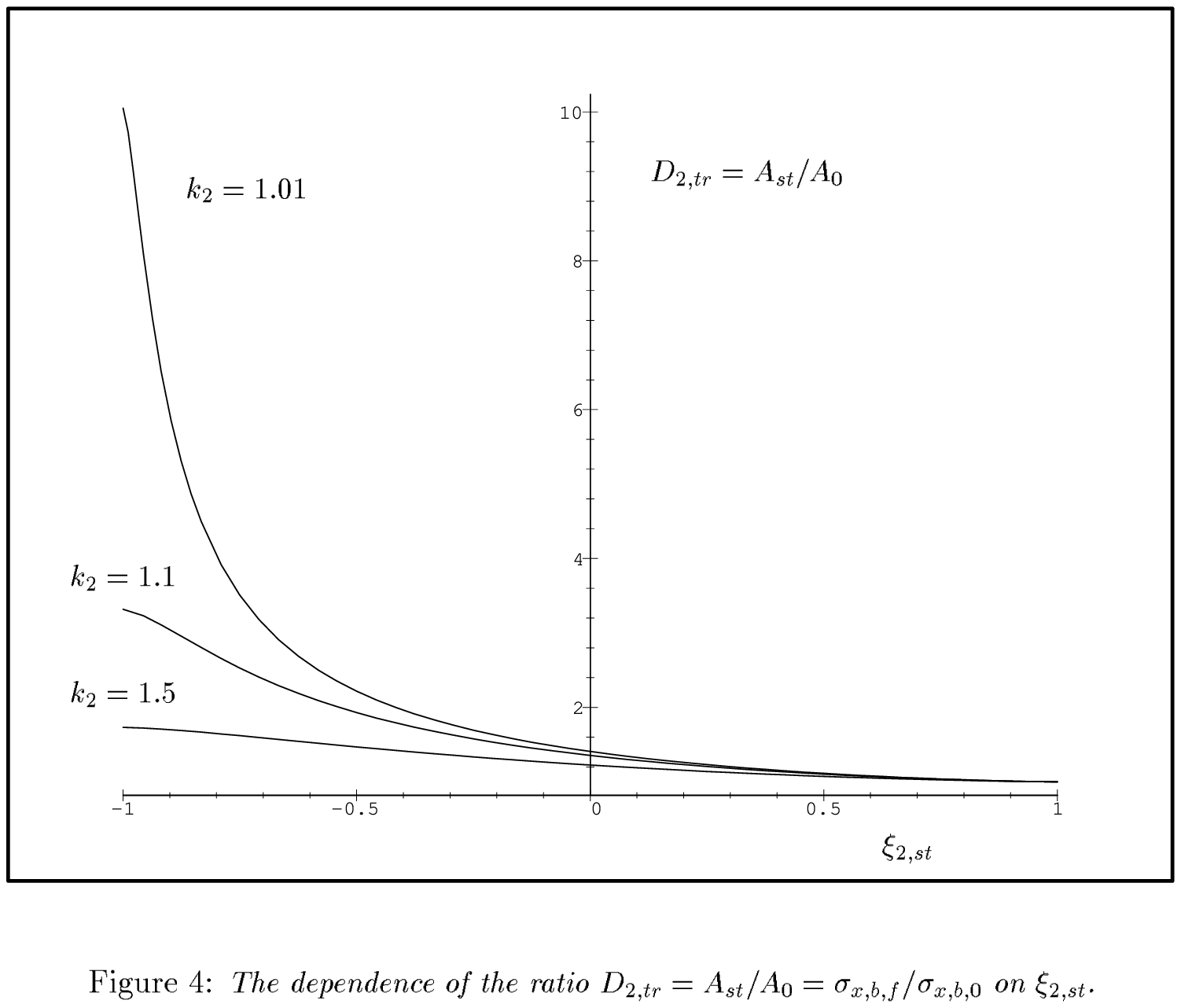}}
\end{figure}

First the target $T _2$ interacts with instantaneous orbits of
particles $x _{\eta} ^{'}$ having maximum energy and amplitudes of
betatron oscillations. The energy of particles is decreased and
amplitudes of betatron oscillations are increased. The instantaneous
orbits of particles $x _{\eta} ^{"}$ having zero amplitudes and the
same maximum energy will be at rest up to a moment $t _0 ^{'}$. The
difference between positions of these orbits $\Delta x _{\eta, 1}$ at
the moment $t _0 ^{'}$ will be maximum by the value and negative.
Instantaneous orbits of particles having maximum initial energies and
total set of amplitudes of betatron oscillations at this moment will be
mixed with instantaneous orbits of particles having lesser initial
amplitudes and energies. The maximum value $|\Delta x _{\eta, 1}|
_{max} \leq 0.28 \sigma _{x,b,0}$. After the moment $t _0 ^{'}$ the
distance $|\Delta x _{\eta, 2}(t)|$ between the orbits $x _{\eta} ^{'}$
and $x _{\eta} ^{"}$ will be decreased, at some moment $t _0 ^{"}$ will
be equal zero $\Delta x _{\eta} ^{"} = 0$, and then it will be
increased up to the moment $t _c$ ($A _{T _2} = l _c$), where it will
be maximum. After the moment $ t _c$ both the amplitude of betatron
oscillations and the spread of instantaneous orbits $\Delta x
_{\eta,2}$ will be constant. This assertion is valid for the
instantaneous orbits of the arbitrary energy.

When the value $\Delta x _{\eta,1}$ is negative then the instantaneous
orbits of particles having minimum initial energies and maximum
amplitudes of betatron oscillations will leave behind the instantaneous
orbits of particles having the same initial energies and zero
amplitudes of betatron oscillations. When the value $\Delta x
_{\eta,2}$ is positive then the instantaneous orbits of particles
having maximum initial energies and amplitudes of betatron oscillations
will fall behind from the instantaneous orbits of particles having the
same initial energies and zero amplitudes of betatron oscillations.
That is why the energy spread of the beam (3.5) will have an addition
$\Delta \sigma _{x,\varepsilon} = |\Delta x _{\eta,1}| + 0.5[\Delta x
_{\eta,2} + |\Delta x _{\eta,2}|]$. This addition is proportional to
the initial spread of amplitudes of betatron oscillations of particles
in the beam.

      \begin{center} \bf Examples \end{center}

In the Tables 8, 9 the examples are presented for the longitudinal
cooling of ion beams at $k _2 = 1.0$ and $k _2 = 1.1$. The next
notations were used: $A _{T_2}/\sigma _{x,b,0} = \pi \psi (k _2, \xi
_{2,st})$; $\sigma _{x,\varepsilon,0} = A _{T _2} - \sigma _{x,b,0}$,
$D _{2,tr} = \sigma _{x,b,f}/ \sigma _{x,b,0}$, $C _{2,l} = \sigma
_{x,\varepsilon,0}/ \sigma _{x,\varepsilon,f}$; $C _{x,b} = \sigma
_{x,0}/ \sigma _{x,f}$.

\begin{figure}[hbt]
\vskip 4mm
Table 8  \hskip 10mm $k _2 = 1.0$, \hskip 5mm $l _c = \infty$
\hskip 5mm $A _{T _2} = \sigma _{x,0} = \sigma _{x,\varepsilon,0} +
\sigma _{x,b,0}$
\vskip 5mm
\begin{tabular}{|l|l|l|l|l|l|l|l|}
\hline
$\xi _{2,st}$ &$\psi(1.0,\xi _{2,st})$&$A _{T _2}/
\sigma _{x,b,0}$&$\sigma _{x,\varepsilon, 0}/\sigma
_{x,b,0}$&$ D_{2,tr}$&$\sigma _{x,\varepsilon, f}/\sigma
_{x,b,0}$ &$C _{2,l}$ &$C _{x,b}$\\
\hline $0.0 $ &0.492&1.545&0.545&1.414& 0.278 &1.96&0.92\\
\hline $-0.2$ &0.716&2.25&1.25&1.636 &0.61    &2.05&1.00 \\
\hline $-0.5$ &1.392&4.38&3.38&2.261 &1.41    &2.40&1.19\\
\hline $-0.9$ &10.187&32.00&31.00&7.314&6.86  &4.52&2.26\\
\hline
\end{tabular} \end{figure}

\begin{figure}[hbt]
\vskip 4mm
Table 9  \hskip 10mm $k _2 = 1.1$, \hskip 5mm $l _c = 12.82\sigma
_{x,b,0}$, \hskip 5mm  $A _{T _2} = \sigma _{x,0} = \sigma _{x,
\varepsilon,0} + \sigma _{x,b,0}$
\vskip 5mm
\begin{tabular}{|l|l|l|l|l|l|l|l|}
\hline $\xi _{2,st}$ &$\psi(1.1,\xi _{2,st})$&$A _{T _2}/ \sigma
_{x,b,0}$&$\sigma _{x,\varepsilon, 0}/\sigma _{x,b,0}$&$ D _{2,tr}
$&$\sigma _{x,\varepsilon, f}/\sigma _{x,\varepsilon,0}$ &$C
_{2,l}$ &$C _{x,b}$\\
\hline $0.0 $ &0.423&1.33&0.33&1.35&0.28&$1.17$&0.82\\
\hline $-0.2$ &0.595&1.87&0.87&1.52&0.584&1.49 &0.89 \\
\hline $-0.5$ &1.033&3.25&2.25&1.93&1.245&1.80 &1.02\\
\hline $-0.9$ &2.759&9.54&8.54&3.04&3.02 &2.82 &1.70 \\
\hline $-1.0$ &3.71&12.82&11.82&3.32&3.60&3.06 &1.85 \\
\hline $-1.0$ &3.71&101  &100  &3.32&11.61&8.61 &6.76 \\
\hline
\end{tabular}
\end{figure}

The high degree enhanced cooling of a particle beam in the longitudinal
plane and much lesser degree of heating in the transverse one takes
place at the second stage of cooling.

The MAPLE V computer program was used to calculate the dependencies
presented in the Tables and Figures of this Appendix.


\begin{thebibliography}{9}
\bibitem{bohm}
D.Bohm and L.Foldy, Phys. Rev. v.70, 249, (1946). 
\bibitem{sands} 
M.Sands, Phys. Rev., v.97, p. 740 (1955). 
\bibitem{robinson} 
K.W.Robinson, Phys. Rev., 1958, v.111, No 2, p.373;
A.Hoffman, R.Little, J.M.Peterson et al., Proc. VI Int. Conf. High
Energy Accel. Cambridge (Mass.), 1967, p.123.
\bibitem{kolom-leb}
A.A.Kolomensky and  A.N.Lebedev,  Theory  of Cyclic Accelerators. North
Holland Publ., $C^o$, 1966; M.Sands, "The physics of electron storage
rings, an Introduction", SLAC Report 121, Nov. 1970
(unpublished); H.Bruk, Accelerateurs Circulaires de Particules (Press
Universitaires de France, 1966); H.Wiedemann, Particle Accelerator
Physics I \& II (Springer-Verlag, New York, 1993.  
\bibitem{budker} G.I.Budker, Atomnaya Energia 22, 346
(1967); Ya.A.Derbenev, A.N.Skrinskii, Sov. Phys. Rev., v.1, p.165,
1981.  
\bibitem{wineland} D.J.Wineland and H.Dehmelt, Bull. Am.
Phys. Soc.  20, 637, (1975); T.W.H{$\ddot a$}nsch and A.L.Shawlov, Opt.
Commun., v.13, p.68 (1975).  
\bibitem{channel} P.J.Channel,
J.Appl.Phys., v. 52(6), p.3791 (1981); L.D.Selvo, R.Bonifacio,
W.Barletta, Optics Communications, v.116, (1995), p.374.  
\bibitem{shroder}
S.Shr{$\ddot o$}der, R.Clein, N.Boos et al., Phys. Rev. Lett., v. 64,
No 24, p.2901 (1990). 
\bibitem{hangst1}
J.S.Hangst, M.Kristensen, J.S.Nielsen et al., Phys. Rev. Lett., v. 67.
1238 (1991); J.S.Hangst, K.Berg-Sorensen, P.S.Jessen et al. Proc. IEEE
Part. Accel. Conf., San Francisco, May 6-9, NY, 1991, v.3, p.1764.  
\bibitem{oneil}
O'Neil G., Phys. Rev., 102, 1418 (1956); A.Shoch, Nucl. Instr. Meth,
v.11, p.40 (1961). 
\bibitem{kolom} A.A.Kolomensky, Atomnaya Energia, v.19, No 5, 1965,
p.534.    
\bibitem{ado}
Yu.M.Ado, Atomnaya Energia, v.31, No 1, 1971, p.40.  
\bibitem{skrin}
A.N.Skrinskii, Uspekhi Fiz. Nauk, v.138, No 1, p.3, 1982. 
\bibitem{vanderm}
S. van der Meer, CERN Internal Report CERN/ISR-PO/72-31 (1972).  
\bibitem{rubbia}
a) C.Rubbia, Nucl. Instr. Meth. A278, 253 (1989); b) M.-J.Miesner,
R.Grimm, M.Grieser, et al., Phys. Rev. Lett. v.77, No4, p.623 (1996);
\bibitem{grim}
R.Grimm, U.Eisenbarth, M.Grieser et al., Advanced
ICFA Beam Dynamics Workshop on Quantum Aspects of Beam Physics, World
Scientific, Ed.  Pisin Chen, Monterey, California, USA, 1998, p.200.
\bibitem{lauer}
I.Lauer, U.Eisenbarth, M.Griezer, et al., Phys. Rev. Lett., v.31, No
10, p.2052, 1998.    
\bibitem{bw}
E.G.Bessonov, F.Willeke, Preprint DESY HERA 95-09,
December 1995. 
\bibitem{zel'dovich} 
Ya.B.Zel'dovich, Sov. Phys. Uspekhi, v.18, No 2, p.79 (1975).
\bibitem{robinson2}
K.W.Robinson, "Storage Ring for Obtaining Synchrotron Radiation Line
Spectra", CEAL 1032 (1966). 
\bibitem{murphy}               
J.B.Murphy, Microbunches Workshop, AIP Conference Proceedings 367,
Upton, NY, September, 1995, p.109.
\bibitem{hutton}               
A.Hutton, Particle accelerators, 1976, v.7, p.177.
\bibitem{chasman}               
R.Chasman, G.K.Green, Proc. 5th All-Union Conference on charged particle
accelerators, Dubna, 5-7 Oct. 1976, v.2, p.172, Moscow, Nauka, 1977.
\bibitem{emery} 
L.Emery, Proc. Workshop on 4th Generation Light Sources, Febr.24-27,
1992, p.149, Ed.M.Cornacchia and H.Winick, SSRL 92/02.
\bibitem{idea}
E.G.Bessonov, a) Preprint FIAN No 6, 1994; b) Proc. of
the Internat. Linear Accel. Conf. LINAC94, Tsukuba, KEK, August 21-26,
1994, Vol.2, pp.786-788; c) Journal of Russian Laser Research, 15, No 5,
(1994), p.403.   
\bibitem{fpl}
E.G.Bessonov, Proc. of the 15th Int. Free-Electron Laser
Conference FEL94, Nucl. Instr. Meth. v.A358, (1995), pp. 204-207.
\bibitem{prl}
E.G.Bessonov and Kwang-Je Kim, Phys. Rev. Lett., 1996,
vol.76, No 3, p.431.     
\bibitem{pac}
E.G.Bessonov,  K.-J.Kim, Proc. of the 1995 Part.
Accel. Conf. and Int. Conf. on High-Energy Accelerators, p.2895;
Proc. 5th European Particle Accelerator Conference, Sitges, Barcelona,
10-14 June 1996, v.2, p. 1196. 
\bibitem{neuffer}
D.V.Neuffer, Nucl. Instr. Meth., 1994, v.A350, p.24. 
\bibitem{hangst}
J.S.Hangst et al., Phys. Rev. Lett., v.74, No 22, p.4432 (1995). 
\bibitem{kim}
K.-J.Kim, Proc. of the 7th Workshop on Advanced Accelerator Concepts,
Lake Tahoe, CA, 1996, AIP Conference Proceedings 398, p.243;
LBNL-40004 (1996) 
\bibitem{zhirong}
Zh. Huang, R.D.Ruth, Phys. Rev. Lett., v.80, No 5, 1998, p. 976. 
\bibitem{bes78}
E.G.Bessonov, Preprint FIAN No 35, Moscow, 1978; Proc. VI All-Union
Conf. on charged particle accelerators (Dubna, Oct. 11-13, 1978), Dubna
JINR, 1979, p.203; Trends in Physics, 1978: Proc. IV EPS General Conf.,
York (UK), 1978, Bristol 1979, p.471.  
\bibitem{petrich}
W.Petrich et al., Phys. Rev. A, v.48, No 3, 1993, p.2127. 
\bibitem{bond}
E.Bonderup, Proc. 5th Accelerator School, Ed.: S.Turner, CERN 95-06,
v.2, p.731, Geneva, 1995.  
\bibitem{habs}
D.Habs et al., Proc. of the Workshop on Electron Cooling and New
Techniques, Lengardo, Padowa - Italy, 1990, World Scientific, p.122.
\bibitem{besabstr}
E.G.Bessonov, Bulletin of the American Physical
Society, Vol.40, No 3, May 1995, p.1196. 
\bibitem{sessler}
H.Okamoto, A.M.Sessler, and D.M$\ddot o$hl, Phys. Rev.
Lett. {\bf 72}, 3977 (1994); T.Kihara, H.Okamoto, Y.Iwashita, et al.
Phys. Rev. E, v.59, No 3, p. 3594, (1999). 
\bibitem{bkw}
E.G.Bessonov, K.-J Kim, F.Willeke, Physics/9812043.  
\bibitem{blp}
V.B.Berestetskii, E.M.Lifshitz and L.P.Pitaevskii, Quantum
Electrodynamics, 2nd ed. (Pergamon Press, New York, 1982).  
\bibitem{srn}
J.Feldhause, B.Sonntag, Synchrotron Radiation News, 1998, v.1,
No 1, p.14.       
\bibitem{ting}
A.C.Ting, P.A.Sprangle, Particle Accelerators, 1987, v.22, p.149 
\bibitem{dikan}
N.S.Dikanskii, A.A.Mikhailichenko, Preprint 88-9, BINP, Novosibirsk,
1988; The Proc. VI All-Union Particle Accelerator Conf., v.1, p.419,
1988, Dubna D9-89-52, 1989 
\bibitem{spr}
P.Sprangle, E.Esarey, in High Brightness Beams for Advanced
Accelerator Application, (AIP, New-York, 1992), p. 87; Phys.
Fluids B4, 1992, p.2241.  
\bibitem{telnov}
V.Telnov, Phys. Rev. Lett., 78 (1997), p.4757; Advanced ICFA Beam
Dynamics Workshop on Quantum Aspects of Beam Physics, World Scientific,
Ed. Pisin Chen, Monterey, California, USA, 1998, p.173; Proc. of the
Int. Symp. on New Visions in Laser-Beam Interactions, Oct. 11-15, 1999,
Tokyo, Metropolitan University Tokyo, Japan. To be published in Nucl.
Instr. Meth. B (see also arXiv:hep-ex/0001028)  
\bibitem{bes2}
E.G.Bessonov,
Advanced ICFA Beam Dynamics Workshop on Quantum Aspects of Beam
Physics, World Scientific, Ed. Pisin Chen, Monterey, California, USA,
1998, p.330.  
\end{thebibliography}
\end{document}